\documentclass[12pt,a4,notitlepage,fleqn]{article}

\usepackage[verbose]{geometry}
\usepackage[symbol]{footmisc}
\usepackage[T1]{fontenc}
\usepackage{authblk}

\makeatletter

\newcommand{\fboxsubsec}[1]{
	\begin{flushleft}
		#1
	\end{flushleft}
	}
	\newcommand{\fboxsubsubsec}[1]{
	\begin{flushleft}
		#1
	\end{flushleft}
	}
\renewcommand{\subsection}{\@startsection{subsection}{2}{0pt}
	{1ex}
	{0.5ex}
	{\reset@font\it\fboxsubsec}
	}
\renewcommand{\subsubsection}{\@startsection{subsubsection}{2}{0pt}
	{1ex}
	{0.5ex}
	{\reset@font\fboxsubsubsec}
	}
\makeatother

\title{Market Integration in Prewar Japanese Rice Markets}%

\author{Mikio Ito$^{a}$, \ Kiyotaka Maeda$^{b}$ \ and \ Akihiko Noda$^{c}$\thanks{\scriptsize Corresponding Author. E-mail: noda@cc.kyoto-su.ac.jp, Tel: +81-75-705-1510, Fax: +81-75-705-3227.}

{\scriptsize ${}^{a}$ \it Faculty of Economics, Keio University, 2-15-45 Mita, Minato-ku, Tokyo 108-8345, Japan}

{\scriptsize ${}^{b}$ \it Faculty of Economics, Seinan Gakuin University, 6-2-92 Nishijin, Sawara-ku Fukuoka 814-8511, Japan}

{\scriptsize ${}^{c}$ \it Faculty of Economics, Kyoto Sangyo University, Motoyama, Kamigamo, Kita-ku, Kyoto 603-8555, Japan}

{\scriptsize ${}^{d}$ \it Keio Economic Observatory, Keio University, 2-15-45 Mita, Minato-ku, Tokyo 108-8345, Japan}}
\date{This Version: \today}

\renewcommand\thefootnote{\arabic{footnote}}

\usepackage{graphicx} 

\usepackage[]{natbib}%
\usepackage{amsmath,amssymb}%
\usepackage{ascmac}%
\usepackage{multirow}%
\usepackage{lscape}%
\usepackage{subfigmat}

\usepackage{pifont}%
\usepackage{arydshln}%
\usepackage[format=hang]{caption}
\usepackage[all]{xy}
\usepackage{url}
\bibpunct{(}{)}{;}{a}{}{,}

\def\hsymbu#1{\smash{\lower1.7ex\hbox{\huge$#1$}}}

\newcommand{\bm}[1]{\mbox{\boldmath{$#1$}}}

\newcommand{\citetapos}[1]{\citeauthor{#1}'s \citeyearpar{#1}}
\newcommand{\citeapos}[2]{\citeauthor{#1}'s (\citeyear{#2})}

\newcommand{\bs}{\texttt{\symbol{'134}}}
\def\cad#1{\texttt{\bs #1} & \csname #1\endcsname }

\begin{document}

\begin{titlepage}

\renewcommand{\thepage}{}
\renewcommand{\thefootnote}{\fnsymbol{footnote}}

\maketitle

\vspace{-10mm}

\noindent
 \hrulefill

\noindent
{\bfseries Abstract:} This paper examines the integration process of the Japanese major rice markets (Tokyo and Osaka) from 1881 to 1932. Using a non-Bayesian time-varying vector error correction model, we argue that the process strongly depended on the government's policy on the network system of the telegram and telephone; rice traders with an intention to use modern communication tools were usually affected by the changes in policy. We find that (i) the Japanese rice markets had been integrated in the 1910s; (ii) increasing use of telegraphs had accelerated rice market integration from the Meiji period in Japan; and (iii) local telephone system, which reduced the time spent by urban users sending and receiving telegrams, promoted market integration.\\



\noindent
\hrulefill

\end{titlepage}

\pagebreak

\section{Introduction}\label{mi_sec1}
The integration of markets in different locations has attracted many economic historians, who are especially interested in agricultural commodity markets. An empirically significant question is: what promotes the integration process? However important this question is, we should consider the point of view from which we answer it. When agricultural commodity markets are discussed in a historical context, transportation cost plays a crucial role because its existence usually inhibits the law of one price. On the other hand, when modern and worldwide international foreign exchanges are discussed, we usually do not consider transportation cost, and feature the efficient market hypothesis and a vector error correction (VEC) model to examine the extinguishing arbitrages among markets that are geographically and informationally separate (see \citet{fama1970ecm} and \citet{gonzalez2001epd}). In this article, we can capture the degree of market integration by measuring the adjustment speed back to the long-run equilibrium relationship among prices, regardless of how markets are disturbed. According to this approach, the time series of market prices are significant rather than is trading volume. Thus, we adopt a VEC model to estimate the long-run equilibrium relationship and adjustment speed; notice that the latter could reflect some tendency toward no arbitrage situations, even when markets are informationally separated.

Paying attention to historical aspects, there is much extant literature on market integration that has attracted many scholars (see \citet{orouke1997egi}, \citet{findlay2007ppt}, and \citet{obstfeld2004gcm}). \citet{federico2012hmd} argues that these studies tend not to focus on what promoted integration, but they show three factors of integration. The first factor was an improvement of market efficiency in the sense of \citet{fama1970ecm}. \citet{federico2007mim} suggests that an increase in market efficiency caused integration of Italian commodity markets before the early 19th century when steam locomotives and steamships became popular.

The second factor was improvement of new transportation channels. \citet{kaukiainen2001swi} and \citet{gonzalez2012igm} mention that it enabled rapid and expanding commodity transactions with distant locations, and promoted market integration in the early 19th century before telecommunication was introduced. 

The third factor was the introduction of telecommunication network. This factor reduced information barriers and directly accelerated to integrate markets (see, for example, \citet{duboff1983tsm}, \citet{field1992mtp}, and \citet{hoag2006atc}). Furthermore, recent research pays attention to the relationship between the second and third factors. \citet{lew2006tct} examine that the diffusion of telecommunication after the transatlantic telegraph cable in 1865 reduced transportation cost because more and more shipping information rapidly shared among traders shed useless anchorage, and \citet{jacks2006wdc} and \citet{federico2011wem} argue that the growing use of transportation because of the reduction in shipping cost contributed market integration. In brief, the diffusion of telecommunication reduced transportation cost, and enabled rapid and massive trading.

At the same time, this situation exposed traders to greater price risks. Accordingly, futures markets were established from the middle 19th century (see \citet[p.11]{kaufmann1984hfm}). These markets in futures efficiently collected and disseminated vast amounts of information, and improved in market efficiency. That is, the diffusion of telecommunication also served as impetus for the establishment of commodity exchanges. In summary, the diffusion of telegraph network was a fundamental factor of market integration from the early modern period.

However, the previous literature does not investigate how the growing in use of telegraph affected the progress of market integration while \citet{stearns2009gwh} shows that the use of telegraph increased since fiercer competition among communication companies reduced telegraph fees. In fact, it is not facile to detect its effect on market integration since the three factors of market integration interrelated as we mentioned above. In particular, the second and third factors resulted in organized markets in which traders could share information efficiently.

On the other hand, in Japan, a commodity futures market was established before the improvement of transportation and information infrastructure. Japanese brokers began to trade rice futures at the {\it Dojima Kome Kaisho} (i.e.,the Osaka-Dojima rice exchange) in the early 18th century  while brokers in Europe and the US established futures exchanges after the middle of the 19th century (see \citet{suzuki1940hdr} and \citet{schaede1989fft}). That is, the brokers could trade commodity futures in Japan earlier than in Europe and the US. \citet{hamori2001eae}, \citet{wakita2001edr}, \citet{takatsuki2008crt}, and \citet{kakizaka2012sed} focus on the performance of the Osaka-Dojima rice exchange and investigate market efficiency. However, the authors fail to create a consensus of the efficiency of the market at that period. In contrast, previous studies succeeded in building a consensus on the market integration in the Tokugawa period. \citet{miyamoto1988mep} and \citet{takatsuki2012rdr} show that traders in western Japan regarded the price in the Osaka-Dojima rice exchange as the standard price for rice in the early 19th century. In short, the Osaka-Dojima rice exchange provided a fine index of the expected rice price and promoted the integration of the Japanese rice markets in the Tokugawa period. The rice futures markets after the Meiji period succeeded the system and customs of trading from the Osaka-Dojima rice exchange in the Tokugawa period (see \citet{suzuki1972crm}).

The Meiji government was formed in 1868, and another rice exchange was established in Tokyo in 1871 (see \citet[pp.27-36]{tge2003htg}). In 1873, the government imported the telegraphic technique from the UK and rapidly built the telegraph network between Tokyo and Osaka, which are almost 250 miles (400 kilometers) apart. That is, the telegraph network was introduced after the start of rice futures trading in Japan. These major rice futures markets were almost efficient until the early 1930s as \citet{ito2016meg} argue. However, \citet{ito2017fpe} focus on the time-varying market efficiency regarding these two markets separately and examines that the difference in time-varying market efficiency among these two markets existed. Nevertheless, there are few studies on the relationship between two major rice markets in Tokyo and Osaka since the previous literature on commodity market integration in Japan focuses on the relationship between the major markets and local markets. \citet{koiwa1974trp,koiwa1980rmp} present that the price trends in the local markets in the 1870s were highly correlated with those in Tokyo. \citet{nakanishi2002cmm} shows that some correlation coefficients between the major and local markets increased from the 1880s to 1900s. These studies do not consider how the growth of telegraph use affected rice market integration while \citet{sugiyama1965svn} suggests that the introduction of information infrastructure promoted market integration. In contrast, the change in the number of telegrams being sent resulted in many revisions of telegram fees (see \citet[pp.117--119]{fujii1998edt}). \citet{mitchener2009icc} show that the diffusion of telegraphs promoted capital market integration among the major and local markets. Accordingly, we detect how the change in the use of the telegraph affected the progress of the integration between two major rice markets.

This paper presents a framework to find the dynamic nature of two major rice futures markets in prewar Japan and study a dynamic aspect of this market integration. The main idea is that the two markets, which worked for their participants to hedge price risk, were co-related and evolving under the development of telecommunications. We consider the time series of both the prices of spot and futures in the two markets. Then, we examine the dynamics of the four dimensional vectors employing a VEC model with time-varying parameters. In particular, considering the possibility of the loading matrix varying over time, we inquire into the dynamic nature of the adjustment process to the long-run equilibrium of the two markets as an aspect of market integration. We propose a measure for the speed of market integration based on the loading matrix and examine its time-varying nature, as well as the historical development of telecommunications in prewar Japan. Our empirical results show that the two major rice futures markets in prewar Japan accelerated their speed of market integration along with the increase in the use of telecommunications.

This paper is organized as follows. Section \ref{mi_sec2} provides a short historical review on rice markets in Japan. Section \ref{mi_sec3} presents our methodology with respect to our genralized least squares (GLS) based technique for a VEC model with time-varying parameters, and provides a statistical inference method for the above time-varying parameters, based on a residual-based bootstrap. Section \ref{mi_sec4} describes the data on the rice futures markets in prewar Japan. Section \ref{mi_sec5} shows our empirical results and their historical interpretation. Finally, Section \ref{mi_sec6} concludes. 

\section{Historical Review of Japanese Rice Markets}\label{mi_sec2}

\begin{figure}[bp]
 \caption{Japanese Cities Related to Rice Distributions}\label{mi_fig7}
 \begin{center}
 \includegraphics[scale=0.8]{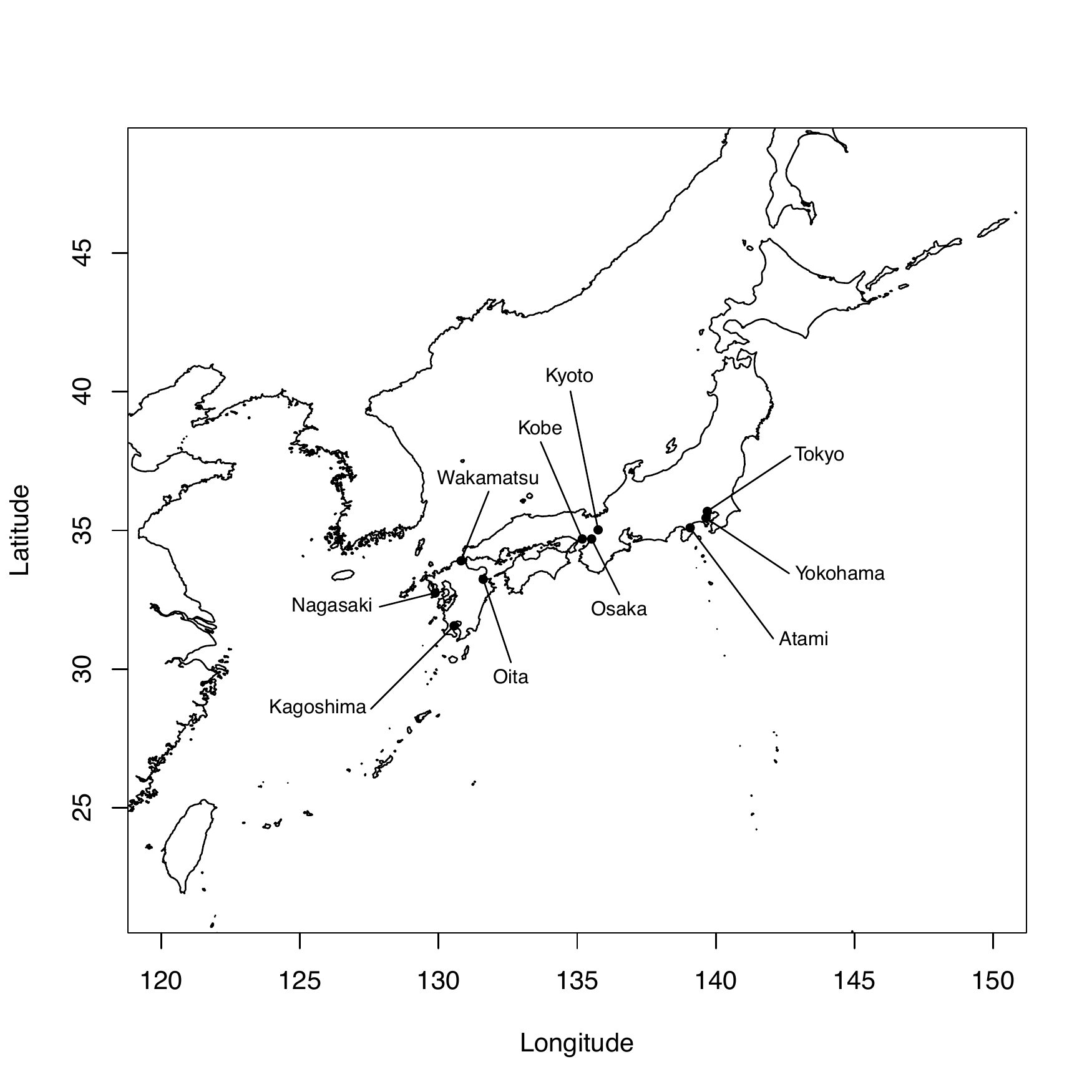}
\vspace*{3pt}
{
\begin{minipage}{350pt}
 \footnotesize
 \underline{Note}: R version 3.4.1 was used to draw the figure.
\end{minipage}}%
\end{center}
\end{figure}

In Japan, rice had been circulated as pseudo money since the 16th century, before the Tokugawa Shogunate was established in 1603. During the period while the Shogunate ruled throughout Japan until 1867, the authority and clan governments (feudal lords) ordered most people to pay their taxes in rice. However, in the early 17th century, the authority began its mintage. In the 1660s, currency was circulated all over Japan; that is, Japan's monetary economy was advancing. At the same time, because the authorities were urged to pay merchants using specie money, they needed to turn their collected rice into money. Accordingly, they sold their rice to brokers in Osaka, which was the major trading center in Japan. Figure \ref{mi_fig7} shows the Japanese cities related to rice distribution.
\begin{center}
(Figure \ref{mi_fig7} around here)
\end{center}
However, because every year the rice crop strongly depended on the climate condition, the rice price fluctuated wildly according to its harvest. Both authorities and brokers faced the volatile price risks whenever they traded rice.

In the early 18th century, brokers began to trade in futures in Osaka to hedge the price risks (see \citet{schaede1989fft}). Then, the Shogunate certified the {\it{Dojima Kome Kaisho}} (i.e., the Osaka-Dojima rice exchange) in 1730. The special rice-price messengers transferred the price at the Osaka-Dojima rice exchange to various local cities in Japan. In addition, rice traders desiring the latest price information used pigeon post and flag signaling. Although these means of communication were prohibited by the Shogunate, they provided the fastest manner of communication at that time (see \citet[pp.335--340]{takatsuki2012rdr}). As a result, the Osaka-Dojima rice exchange provided a standard price for at least western Japan (see \citet[pp.402--403]{miyamoto1988mep}).

In 1868, the Meiji government declared a new regime. Since then, the rice futures market changed. In the 1870s, new rice exchanges were established in Tokyo. In 1871, Hachirouemon Mitsui, the representative Japanese merchant in Tokyo, established a company named {\it{Bouekishousha}} to deal in rice futures. In 1874, another company, {\it{Chugai Shougyo Kaisha}} established by the wealthy merchants in Kagoshima, began to deal in rice futures. In 1883, these two rice exchanges were unified and renamed {\it{Tokyo Kome-Shoukaisho}} (i.e., the Tokyo rice exchange) (see \citet[pp.27--36]{tge2003htg}). Therefore, the rice brokers could trade rice futures in Tokyo and Osaka, the two major cities in Japan, after the 1870s; the environment of rice trading had changed.

Since the 1870s, rice traders could use the telegraph network to obtain the latest information regarding rice trading. The government began to build and operate this network using a British company's technique in the early 1870s. In fact, it laid the first telegraphic line between Tokyo and Nagasaki via Osaka in 1873; users could send and receive telegrams between Tokyo and Osaka. In the next decade, rice traders were able to obtain the latest news on price information more rapidly.

In 1889, the Ministry of Communications began to operate a limited telephone service between Tokyo and Atami on a trial basis. In the following year, the government launched a local and long-distance telephone service in Tokyo and Yokohama on a commercial basis. Moreover, in 1893, the Ministry of Communications opened another local telephone service in Osaka, and a new long-distance telephone service between Tokyo and Osaka was available in 1899 (see \citet[pp.453--454]{mct1940htb4}). The series of improvements of telephone network enabled high-speed communication between Tokyo and Osaka.

In particular, rice traders used the telegraph and telephone frequently. Previous literature on the history of information and communications in Japan mentions that stock dealers in Tokyo and Osaka have exchanged information about stock price with each other using the telegraph and telephone since the Meiji period (see \citet[p.192]{iishi1994shi}). Rice traders also began to use communication facilities and exchange price information between Tokyo and Osaka in the Meiji period, since they often arbitraged between Tokyo and Osaka (see \citet[pp.290--291]{shimizu1913sr}). In fact, Chugai Shogyo Shimpo (i.e., Chugai Commercial Newspaper) reported that a major rice trader in Kyoto made deals in both Tokyo and Osaka in October 1893.\footnote{See \citet{css1893sam} for more details.} In this situation, rice traders obtained the latest trade information for rice, and the rice price in certain markets responded sensitively to price fluctuations in the other markets.

Especially, rice traders in Tokyo and Osaka mainly used the telegram for long distance communications, which was the result of two factors. First, the telegram fee between Tokyo and Osaka was lower than the corresponding telephone fee, as will be described in Section \ref{mi_sec5}. Second, using the long-distance telephone service did not always accelerate the communication speed. The users of long-distance telephone service had to wait a long time for telephone switching because the capacity of the telephone lines was insufficient. In fact, the average waiting time for a telephone call from Tokyo to Osaka was 203 minutes in 1937 (see \citet[p.482]{ntt1960htt}). Consequently, rice traders used the telegram frequently, and they attempted to reduce the characters in the telegram since their fee was charged per number of characters. Accordingly, rice traders sent telegrams in code, which also contributed to the guarding of the secrecy of communications.

\begin{landscape}
\begin{figure}[bp]
 \caption{Telegram Code: A Case of Masuzo Hirai}\label{mi_fig8}
 \begin{center}
 \includegraphics[scale=0.45]{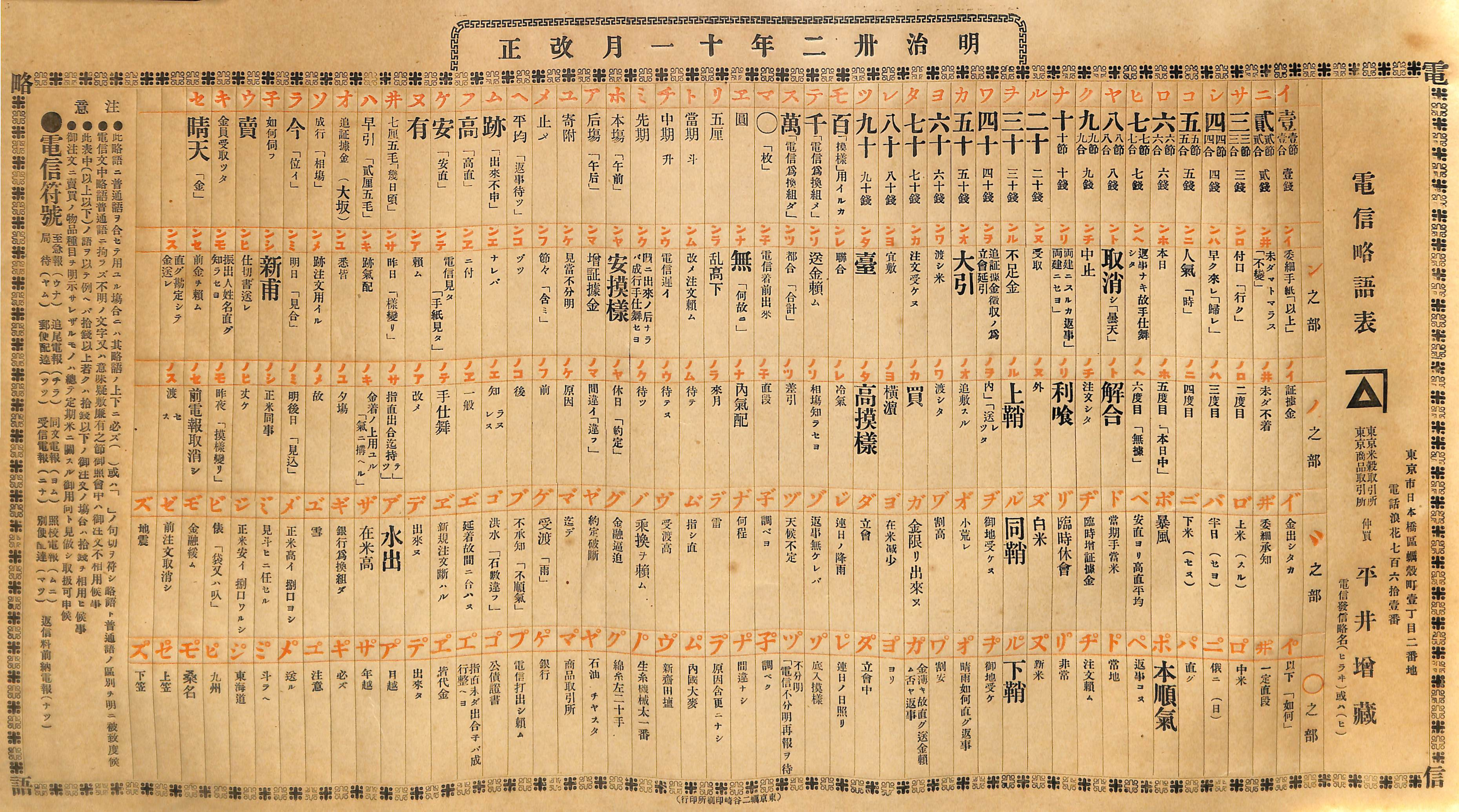}
 \end{center}
\end{figure}
\end{landscape}

Figure \ref{mi_fig8} shows a telegram code that Masuzo Hirai, a representative rice trader in Tokyo, used in 1899; to transmit rice-trading information, many rice traders used their own telegram codes, similar to that in \ref{mi_fig8}, which reduced their telegram fee. In addition, on September 25, 1893, according to Yomiuri Shimbun (i.e., Yomiuri Newspaper), the rice price in Tokyo increased as soon as rice traders there obtained information regarding the large transactions by major rice traders in Osaka.\footnote{See \citet{yomiuri1893yrm} for details.}  Furthermore, the telegram transmitted not only trade information, but also disaster information. For example, in October 1893, when the typhoon passed through western Japan, the telegraph transmitted information on the flood damage in western Japan to Tokyo and Osaka, and the rice price increased in both cities.\footnote{See \citet{css1893irp} for details.}

\begin{center}
(Figure \ref{mi_fig8} around here)
\end{center}

At the same time, rice traders could use another instrument to obtain information. In the 1870s, many publishers of newspapers and business magazines were established; they introduced steam-driven printing machines enabling printing in large quantities (see \citet[pp.31--32]{minami1976mpt}). The publishers sold their newspapers and magazines with various price information acquired by using the telegram for lead articles. They published in large volumes and extensively diffused price information among market participants.

During this same period, the transportation infrastructure was also developed. The Japanese government began to build a railroad network in 1870, and a first mainline, the Tokaido Line, between Tokyo and Kobe via Osaka was opened in 1889 (see \citet[p.42]{muramatsu1965hig}). However, the railroad did not connect cities and farming villages, since it was built between major cities until the end of the 19th century. Consequently, rice had been carried mainly by sailing ships until the late 1890s. In other words, rice traders did not use steam ships for rice transportation, which had the advantages of safety and promptness when compared to sailing ships. In the meantime, sailing ships had the advantage of lower transportation costs, when compared to steam ships, since they did not need fuel. For example, the transportation cost of a sailing ship was about 30\% cheaper than that of a steam ship in 1879 (see \citet[pp.78--79]{morita1975dr}). However, the use of steam ships expanded rapidly in the 1900s.

In 1902, the tonnage of a steam ship exceeded that of a sailing ship in Japan (see \citet[p.346]{miwa1975mt}). Moreover, the widespread use of steam ships improved transportation conditions. First, transportation costs were reduced with the intensifying competition among steam ships. In fact, the cost for coal transportation from the port of Wakamatsu to port of Yokohama, the major index of ship-freight cost in prewar Japan, decreased by 32\% from 1899 to 1903 (see \citet[p.624]{toyo1927jss}).\footnote{The port of Wakamatsu was the major coal port, since Wakamatsu City was near Chikuho coalfield, one of the largest coalfields in Japan.} Second, numerous steamer coastal lines were introduced, and reduced transportation costs. For example, rice traders in Kagoshima paid 80 yen for the transportation of 100 koku of rice to Tokyo in 1897.\footnote{{\it Koku} is a unit of rice trading volume in Japan. One {\it koku} is equal to 180.39 liters.} This cost included the transshipment charge at the port of Kobe, since the direct ship course between Kagoshima and Tokyo was yet not introduced. However, after the introduction of a direct course between both cities, rice traders in Kagoshima paid only 40 yen for the rice freight cost to Tokyo in 1902 (see \citet[pp.456--457]{sasaki1937htr}). In the same year, the annual average rice price per 100 koku in Tokyo was 1,267 yen (see \citet[p.390]{nakazawa1933nbh}). In short, the transportation cost between both cities was only 3\% of the rice price. Even rice traders who were located in remote places, away from major markets, such as Tokyo and Osaka, could use this low-cost transportation in the 1900s.

In the 1910s, the rice price increased in association with the industrialization and urbanization in Japan. From 1910 to 1920, the annual average rice price per 100 {\it koku} in Tokyo increased from 1,303 to 4,428 yen (see \citet[pp.422, 462]{nakazawa1933nbh}). At the same time, World War I began, and the ship freight cost dramatically increased; the ship freight cost to transport coal between Wakamatsu and Yokohama increased by 10.8 times from 1910 to 1918 (see \citet[p.624]{toyo1927jss}). This situation shrunk the cost gap between ship freight and rail transport. From 1890 to 1910, the distance of the railroad in Japan increased from 1,750 to 5,606 miles, and it connected cities and farming villages (see \citet[p.619]{toyo1927jss}). Furthermore, the Japanese government nationalized private mainlines in 1906, and integrated the system of rail transportation. However, the rail transportation cost was still higher than that of ship freight before World War I. For example, the cost to transport 100 koku of rice from Oita to Osaka was 90 yen by rail and 50 yen by ship in 1914; it was 76 yen by rail and 70 yen by ship in 1919. This cost gap shrunk during World War I (see \citet[p.550]{tbmr1925esr}). In the same year, the annual average rice price per 100 {\it koku} in Tokyo was 4,503 yen (see \citet[p.458]{nakazawa1933nbh}). In short, although the ship freight cost increased, the cost for long-distance transportation remained below 2\% of the rice price. After the war, rice traders mainly used the railroad to transport rice. In fact, 98\% of the rice that arrived in Tokyo in 1921 was transported by rail (see \citet[p.273]{sasaki1937htr}).

From the late 19th to early 20th century, when the communication infrastructure was developed, the transportation infrastructure was also rapidly developing. This significantly reduced the transportation cost; however, the rice transportation cost had maintained a small percent of the rice price since the turn of the 20th century. These government-operated networks of telegraph and railroad not only increased the amount of information communication and transportation, but also accelerated their speed. Whereas private firms mainly built telegraph and railroad networks in Europe, the government imported Western techniques in order to rapidly introduce both networks in Japan. Consequently, the Japanese government operated transportation and telecommunication infrastructure on its initiative, and tremendously increased the speed of transportation and information communication between Tokyo and Osaka.

\section{The Model}\label{mi_sec3}
This section presents our framework to estimate a time-varying VEC model and conduct inference for the estimates. Featuring the time series of both the prices of spot and futures in the two markets, we examine the dynamics of the four dimensional vectors employing a VEC model with time-varying parameters. In particular, we focus on the time-varying loading matrix of the model, as it provides information for an aspect of market integration.

\subsection{Framework}
We consider a time-varying VEC model for the four-dimensional, multiple time-series data of spot and futures prices in Tokyo and Osaka:\footnote{See the online appendix of \citet{ito2016tvc} for a more detailed discussion, available at \url{http://at-noda.com/appendix/efficiency_integration_appendix.pdf}.}
 \[
  X_t =
  \begin{pmatrix}
   x_{1t}\\
   x_{2t}\\
   x_{3t}\\
   x_{4t} 
  \end{pmatrix}, (t=1,\cdots,T), 
 \]
 where $x_{1t}$, $x_{2t}$, $x_{3t}$, and $x_{4t}$ denote the logged spot and futures prices of the Tokyo and Osaka rice exchanges, respectively. Furthermore, supposing that there is an appropriate size of data for priors, we obtain $\Delta X_t = X_t - X_{t-1}$. In this paper, we suppose that a long-run relationship, represented by simultaneous linear equations among the above four variables at each period, has drifts and is independent of the time trend. Thus, the VEC model in this paper is represented as follows, in which all coefficients are time-invariant:
\begin{multline}\label{VECM eq Matrix form : const}
\begin{bmatrix}
 \Delta X_1 & \cdots & \Delta X_T
\end{bmatrix}
=
\begin{bmatrix}
 \Gamma_1 & \cdots & \Gamma_k 
\end{bmatrix}
\begin{bmatrix}
 \Delta X_0 & \cdots & \Delta X_{T-1} \\
 \vdots     & \ddots & \vdots \\
 \Delta X_{-k+1} & \cdots & \Delta X_{T-k+1} \\
\end{bmatrix}
 \\
+
\Pi
\begin{bmatrix}
1 & \cdots & 1 \\
X_{1-k} & \cdots &  X_{T-k} \\
\end{bmatrix}
+
\begin{bmatrix}
 \boldsymbol{\varepsilon}_1 & \cdots & \boldsymbol{\varepsilon}_T
\end{bmatrix},
\end{multline}
where ${\boldsymbol{\varepsilon}}_t$ is an exogenous shock in each period following intertemporally independent distributions. The coefficient matrix $\Pi$ has a decomposition of $\boldsymbol\alpha$ and $\boldsymbol\beta$ of their ranks of $r$ that $\Pi=\boldsymbol\alpha\boldsymbol\beta'$, if the cointegration order is $r$. Note that this $\Pi$ is a $4\times 5$ matrix, since we consider that the long-run equations between two rice markets in Tokyo and Osaka include drift terms.
 
\subsection{Least Square Technique for Ordinary Vector Error Correction Model}
ecause a VEC model is algebraically derived from a certain vector autoregressive (VAR) model, a linear stochastic system, it is also such a system. Thus, we can estimate parameters $\boldsymbol\mu$, $\Gamma_i$, and $\Pi$ using some regression techniques, such as OLS or GLS. Let $Z_{0}$, $Z_{1}$, and $Z_{k}$ denote appropriate data matrices, represented in Equation (\ref{VECM eq Matrix form : const}). Furthermore, $\boldsymbol{\varepsilon}$ denotes a matrix of exogenous shock vectors. With regard to Equation (\ref{VECM eq Matrix form : const}), its expression is as follows:
\begin{equation}\label{VECM simple form: c}
 Z_0' = \Gamma Z_{1}' + \Pi_{} Z_{k}' + \boldsymbol{\varepsilon},
\end{equation}
where:
\[
Z_{0}'=
\begin{bmatrix}
 \Delta X_1 & \cdots & \Delta X_{T} 
\end{bmatrix},
\]
\[
Z_{1}'=
\begin{bmatrix}
 \Delta X_0 & \cdots & \Delta X_{T-1} \\
 \vdots     & \ddots & \vdots \\
 \Delta X_{-k+1} & \cdots & \Delta X_{T-k+1} \\
\end{bmatrix},
\]
\[
Z_{k}'=
 \begin{bmatrix}
  1       & \cdots & 1 \\
  X_{1-k} & \cdots &  X_{T-k} \\
 \end{bmatrix},
\]
and
\[
 \Gamma=
 \begin{bmatrix}
 \Gamma_1 & \cdots & \Gamma_k 
\end{bmatrix}.
\]

The matrix $\Gamma$ might include $\boldsymbol\mu$ as well as $\Gamma_1, \cdots, \Gamma_k.$ It provides us with information about the stationary aspect of the time series $X_t.$ On the other hand, the matrix takes a crucial role in a VEC model; it is decomposed into {\it loading matrix} $\boldsymbol\alpha$ and the {\it cointegration matrix} $\boldsymbol\beta$, such that $\Pi=\boldsymbol\alpha\boldsymbol\beta'$. Both matrices $\boldsymbol\alpha$ and $\boldsymbol\beta$ have $r$ values less than four, called the cointegrated rank order. In particular, $\boldsymbol\beta'Z'_k$ signifies some long-run relationship among the variables, represented by equations with values less than four. In this paper, we are interested in the time-varying nature of loading matrix $\boldsymbol\alpha,$ given the long-run relationships. 

One can select the lag order $k$ using the usual information criteria, such as \citetapos{schwarz1978edm} Bayesian information criterion for each linear model above; they are easy to compute. The cointegration order $r$ is usually selected through the well-known Johansen's test; this procedure on the rank helps us to obtain all of the estimates and statistics for applied econometrics (see \citet{johansen1988sac,johansen1991eht}).

\subsection{Time-Varying Vector Error Correction Model}
Since we regard a VEC model as a simultaneous linear regression, as in Equation (\ref{VECM simple form: c}), we can examine the possibility of parameter constancy on $\Pi$ and $\Gamma$ using \citeapos{hansen1992a}{hansen1992a,hansen1992b} parameter constancy test. Its null hypothesis is that parameters are time-invariant, and the alternative hypothesis is that they are martingales.

There are several stochastic processes that are martingales. Thus, when the null hypothesis is rejected, we have to choose one of such processes that would be followed by the time-varying parameters in our model. Because we are interested in gradual changes in the speed of adjustment to a certain long-run equilibrium for a VEC model, we choose a parameter dynamic in which the parameters follow a random walk.

We present our method to estimate the time-varying loading matrix, regarding a VEC model as a simple linear regression, as in Equation (\ref{VECM eq Matrix form : const}). To this end, we can simply apply \citetapos{ito2017aae} method to obtain GLS estimates of the time-varying parameters. In practice, we estimate $\Gamma$ and $\boldsymbol\alpha$, regarding $\boldsymbol\beta$ as time-invariant and given.
\begin{equation}\label{parameter dynamics Gamma}
\Gamma_t = \Gamma_{t-1} + \bm{u}_{t}, 
\end{equation}
\begin{equation}\label{parameter dynamics alpha}
 \boldsymbol\alpha_t = \boldsymbol\alpha_{t-1} + \bm{v}_{t}. 
\end{equation}
First, we build an $r$ dimensional time series, $Y=Z_{k}\boldsymbol\beta$, where $r$ is the cointegrating order. Then, we apply \citetapos{ito2014ism} method again to a new time linear regression:
\begin{equation}\label{VECM linear model wtih constant beta}
  Z_{0t}' = \Gamma_t Z_{1t}' + \boldsymbol\alpha_t Y_t' + \boldsymbol\varepsilon_t,
\end{equation}
where $Z_{0t}$, $Z_{1t}$, $Y_{t}$, and $\boldsymbol\varepsilon_t$ are the $k$th columns of  $Z_{0}$, $Z_{1}$, $Y$, and $\boldsymbol\varepsilon$, respectively, considering Equations (\ref{parameter dynamics Gamma}) and (\ref{parameter dynamics alpha}). Note that both $\boldsymbol\alpha_t$ and $\boldsymbol\beta_t$ for each $t$ cannot be estimated. In particular, since $\Pi_t = \boldsymbol\alpha_t \boldsymbol\beta'_t$ for each $t$ and $\Pi_t$ is not of full rank, a decomposition of $\Pi_t$ into $\boldsymbol\alpha$ and $\boldsymbol\beta$ is not unique. Thus, either $\boldsymbol\alpha_t$ or $\boldsymbol\beta_t$ is supposed time-invariant for the most general case of  both $\Pi_t$ and $\Gamma_t$ supposed time-varying. We regard $\boldsymbol\beta'Z'_k=\bm{0}$ as long-run equilibrium relations with respect to the multiple time series; we regard $\boldsymbol\beta$ as a constant matrix. In order to confirm the constancy of $\boldsymbol\beta$, we conduct the parameter constancy test of \citet{hansen1999stp}. 

\subsection{Speed of Market Integration}
Under the constancy of $\boldsymbol\beta$, we can regard the loading matrix $\boldsymbol\alpha$, representing the speed of adjustment, as time-varying, when we use a time-varying VEC model. Thus, we focus our attention on the time-varying loading matrix $\boldsymbol\alpha_t$, which provides information about the dynamics of market integration. The larger the absolute value of its components, the more significant their contribution to ameliorate deviation from the long-run equilibrium. Thus, we propose an index of market integration based on the loading matrix. Then, we apply the index for the time-varying loading matrix $\boldsymbol\alpha_t$ to investigate how the speed of market integration varies.

We derive the index $\zeta_t$ from $\boldsymbol\alpha_t$ following \citet{ito2014ism}. In particular:
\begin{equation}\label{eqn index}
 \zeta_t = \sqrt{\max \lambda(\boldsymbol\alpha_t \boldsymbol\alpha_t')}.
\end{equation}
That is, $\zeta_t$ is the square root of the largest eigen value of $\boldsymbol\alpha_t \boldsymbol\alpha_t'$, which is a non-negative semidefinite matrix, for each $t$. Notice that the higher the index, the faster the adjustment of markets to the long-run equilibrium. 

\subsection{Confirming Our Assumption of Constant Cointegrating Vectors}
We make several substantial assumptions on our time-varying VEC model. Among these, the assumption that the cointegrating vectors are constant over time, while the loading matrix is time-varying, is crucial. When we estimate the time-varying coefficients, we need to confirm that the cointegrating vectors are stable. In fact, one could say that if the matrix $\Pi$ is time-varying, the number of cointegrating vectors (as well as the cointegrating vectors themselves) is time-varying. Then, we apply a fine test proposed by \citet{qu2007scd} to our rice data in order to confirm the stable cointegrating vectors.\footnote{As detailed in \citet{juselius2006cvm}, there are a number of econometrics tests that investigate the constancy of cointegrating vectors. In our view, however, the \citet{qu2007scd} test is best suited for our purpose because it considers a variety of test statistics and their asymptotic properties, without assuming the timing of possible structural breaks.} Essentially, by using this test, we can assess whether the number of cointegrating vectors has changed during a subsample, such as the time between $T_{a}$ and $T_{b}$, where neither $T_{a}$ nor $T_{b}$ is known to the researcher. The null hypothesis of \citetapos{qu2007scd} test states that the number of cointegrating vectors is constant for the whole sample. Hence, when we cannot reject the null hypothesis, we can say that we have a certain justification of our assumption that the cointegrating vectors are stable over time.

\section{Data}\label{mi_sec4}

We conduct a dataset of monthly rice spot and futures prices in prewar Japan (Tokyo and Osaka), when we argue the integration of the two major rice futures markets, with respect to the telegram and telephone networks of prewar Japan. It would be no problem to use the data for our goal. The argument of financial market integration is mathematically related to the cointegration of the return series of financial commodities; this property is irrelevant to a wide range of transformations. For example, when an original daily series has this property, the monthly series obtained by the monthly averaging of the daily series still has the property of cointegration. Furthermore, another monthly series obtained by selecting the last observation for each month also has this same property. That is, even if we employ lower frequency data due to availability, it would not be a problem, except for reducing the precision for statistical analyses, when we address market integration.

For the rice futures prices, we only utilize weighted average monthly values for the deferred contract (three months) in the Tokyo and Osaka rice exchanges because it is widely known that the deferred contract month transactions amounted to approximately 70\% of all futures in both rice exchanges.\footnote{Figures 3 and 4 of \citet{ito2017fpe} show that the farthest contract month transaction was larger than any other futures transaction in both Tokyo and Osaka.} For the rice spot data, spot prices in Tokyo are on the Tokyo-Fukagawa rice spot market, and the spot prices in Osaka are the wholesale prices of rice in Osaka. These datasets consist of the following three statistics: (i) Nakazawa (1933) for all futures prices from October 1880 through November 1932, (ii) the {\it Tokyo City Statistics} and \citet[p.342]{mci1931swp} for the spot prices in Tokyo from April 1881 through November 1932, and (iii) \citet{miyamoto1979owp} and the {\it Osaka City Statistics} for the spot prices in Osaka from April 1881 through November 1932. There are a few missing values in both sets of statistics; therefore, we fill in the missing values using a seasonal Kalman filter.

For our estimations, we take the natural log of the spot and futures rice prices, to obtain the level data, as well as the first difference of the natural log of the level data, to obtain the returns on spot and futures prices, as the first difference data.

Table \ref{mi_table1} shows the results of the unit root test with descriptive statistics for the data. We confirm whether the variables satisfy the stationarity condition and apply the augmented Dickey-Fuller generalized least squares (ADF-GLS) test of \citet{elliott1996eta}. We employ the modified Bayesian information criterion (MBIC) instead of the modified Akaike information criterion (MAIC) to select the optimal lag length because, from the estimated coefficient of the detrended series, $\hat\psi$, we do not find the possibility of size distortions (see \citet{elliott1996eta}; \citet{ng2001lls}). There is widespread agreement that the logarithmic spot and futures rice prices are both integrated of the order one (or $I(1)$ process), so that the differences are stationary (or $I(0)$ process) variables.
\begin{center}
(Table \ref{mi_table1} around here)
\end{center}

\section{Empirical Results}\label{mi_sec5}
\subsection{Preliminary Results}
We first confirm cointegration relations by using \citetapos{johansen1988sac} maximal eigenvalue test and \citetapos{johansen1991eht} trace test. Table \ref{mi_table2} presents the results of the tests. The eigenvalue and trace statistics are shown in the third and the fifth columns, respectively. Both the results reject the three null hypotheses of ``$r=0$ (no cointegration),'' ``$r=1$ (at most one cointegration),'' and ``$r=2$ (at most two cointegration),'' in favor of ``$r=3$ (at most three cointegrations)'' at the 1\% significance level. This indicates the presence of three cointegrating relationships.
\begin{center}
(Table \ref{mi_table2} around here)
\end{center}

We assume a time-invariant VEC model for our preliminary estimations. Then, we regress the difference of each of the four rice prices on their lagged values and error correction terms of the time-invariant VEC model. Table \ref{mi_table3} summarizes our estimation results for a time-invariant VEC model for the whole sample. Notice that only three error correction terms significantly affect the price changes of spot and futures in Tokyo and Osaka. This is consistent with the above results of Johansen's cointegration tests. In particular, the price change of futures in Tokyo depends only on its lagged value.
\begin{center}
(Table \ref{mi_table3} around here)
\end{center}

Moreover, Table \ref{mi_table3} also shows some asymmetric structure of rice markets in Tokyo and Osaka. First, the change in futures rice prices in Osaka depends not only on its lagged value, but also on the corresponding value in Tokyo. Second, the change in spot rice price in Osaka does not depend on its lagged value, but on the corresponding value in Tokyo. Third, both the changes in spot and futures rice prices in Tokyo solely depend on their own lagged values. These three facts concern the short-run effects on price changes in rice markets. On the other hand, the error correction terms affect three rice prices in Tokyo and Osaka, except for the change in the futures rice price in Tokyo. Apparently, the rice market in Osaka is dependent, whereas that in Tokyo is independent. However, these results are interesting, in that the time-invariant VEC model fails to capture the chronological market integration of the two rice markets in detail.

As discussed in Section \ref{mi_sec2}, the Japanese government had advanced the building of the telegraph network since the 1870s. Then, more and more market participants could enjoy quick communication on the rice price between Tokyo and Osaka. Therefore, we should examine the evolution of the rice market structure across the two cities corresponding to the acceleration of information propagation. As such, Table \ref{mi_table3} shows the result of \citeapos{hansen1992a}{hansen1992a,hansen1992b} parameter constancy test, which suggests that the parameters of our VEC model vary over time.

Now, relaxing the assumption that the parameters of our VEC model are time-invariant, we employ the time-varying parameter model that presumably could better capture the dynamics of the spot and futures prices of rice when we take into account the gradually changing market environment. Therefore, we would obtain a different market nature if we only once applied the time-varying VEC model. However, before we proceed, we should confirm whether the cointegrating relationship is stable, that is, that the cointegrating vector is preserved for the whole sample period. Using \citetapos{qu2007scd} test, we can conclude that the number of cointegrating order among the spot and futures prices of the two rice exchanges in Tokyo and Osaka does not alter over time at the 1\% significance level. These results are presented in Table \ref{mi_table4}. 
\begin{center}
(Table \ref{mi_table4} around here)
\end{center}

In the next subsection, we estimate the time-varying speed index of market integration using the time-varying loading matrix of the TV-VEC model. It provides information on extensive dynamics of rice market integration of Tokyo and Osaka.

\subsection{Time-Varying Speed of Market Integration and Historical Interpretation}
\subsubsection{Growing Telecommunications and Speed of Market Integration}
\begin{figure}[bp]
 \caption{Time-Varying Speed of Market Integration and its Accelaration}\label{mi_fig1}
 \begin{center}
 \includegraphics[scale=0.5]{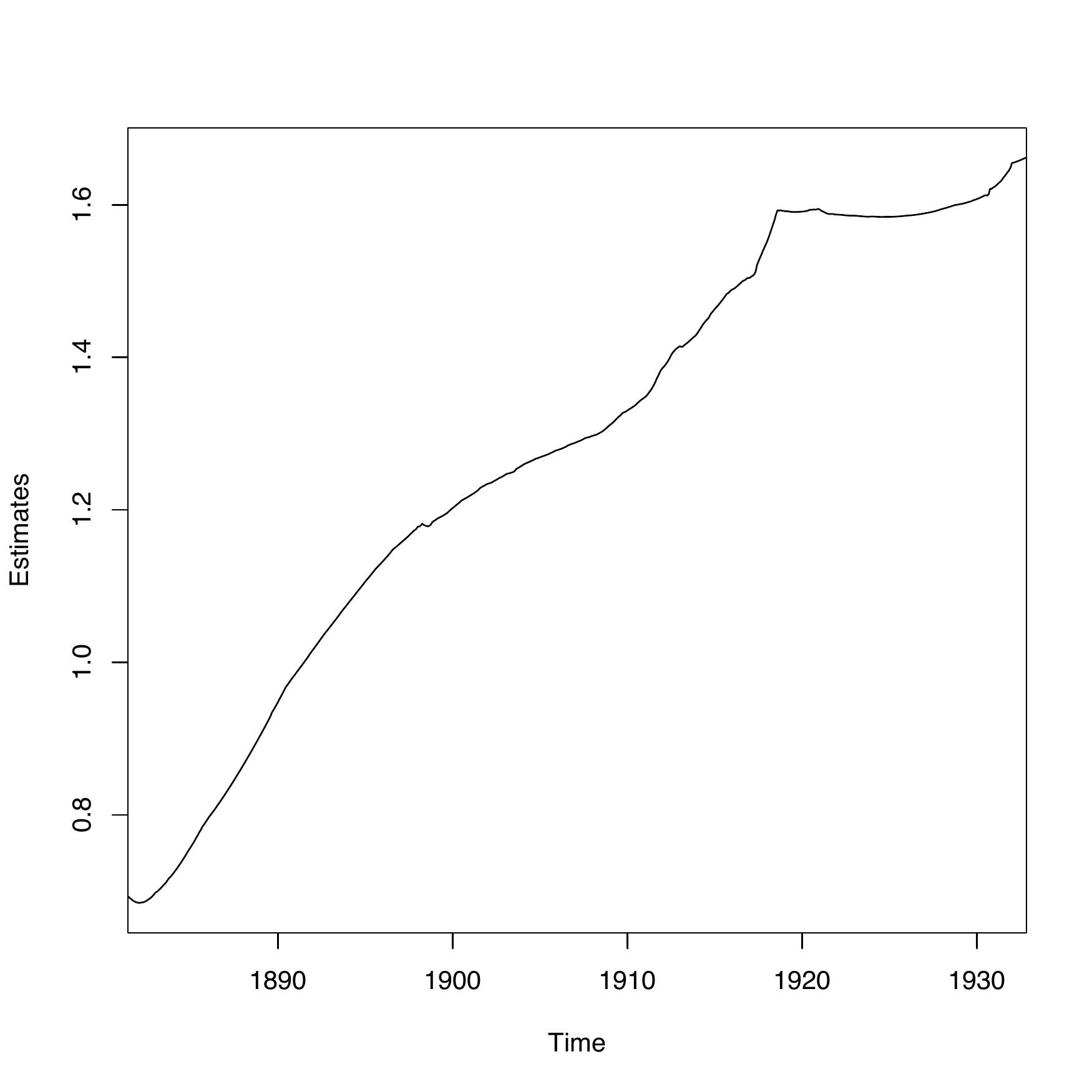}
 \includegraphics[scale=0.5]{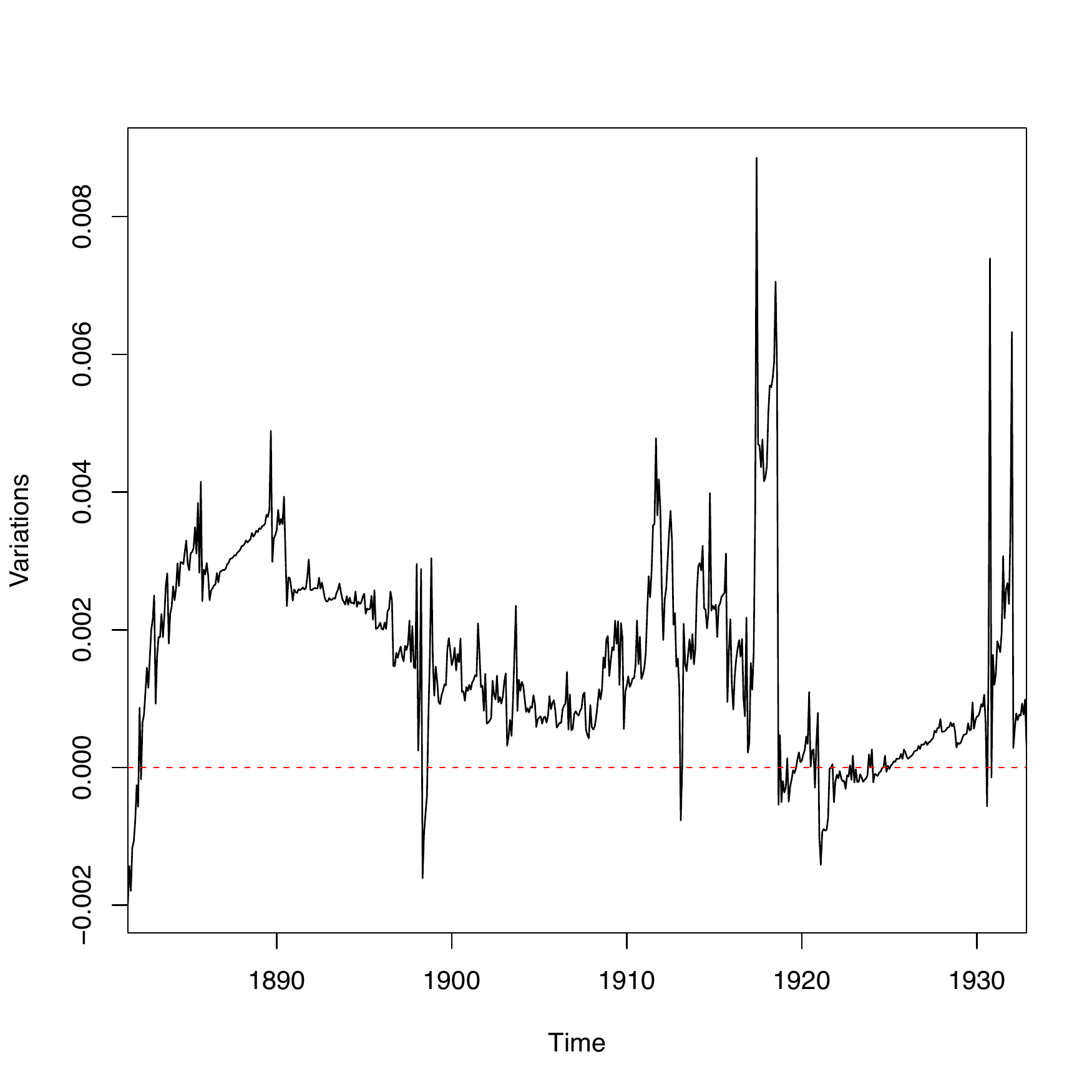}
  \vspace*{3pt}
{
\begin{minipage}{350pt}
\footnotesize
\underline{Note}: R version 3.4.1 was used to compute the estimates and the statistics.
\end{minipage}}%
\end{center}
\end{figure}

\begin{figure}[bp]
 \caption{Real Fees of Telephone and Telegram between Tokyo and Osaka}\label{mi_fig2}
 \begin{center}
 \includegraphics[scale=0.8]{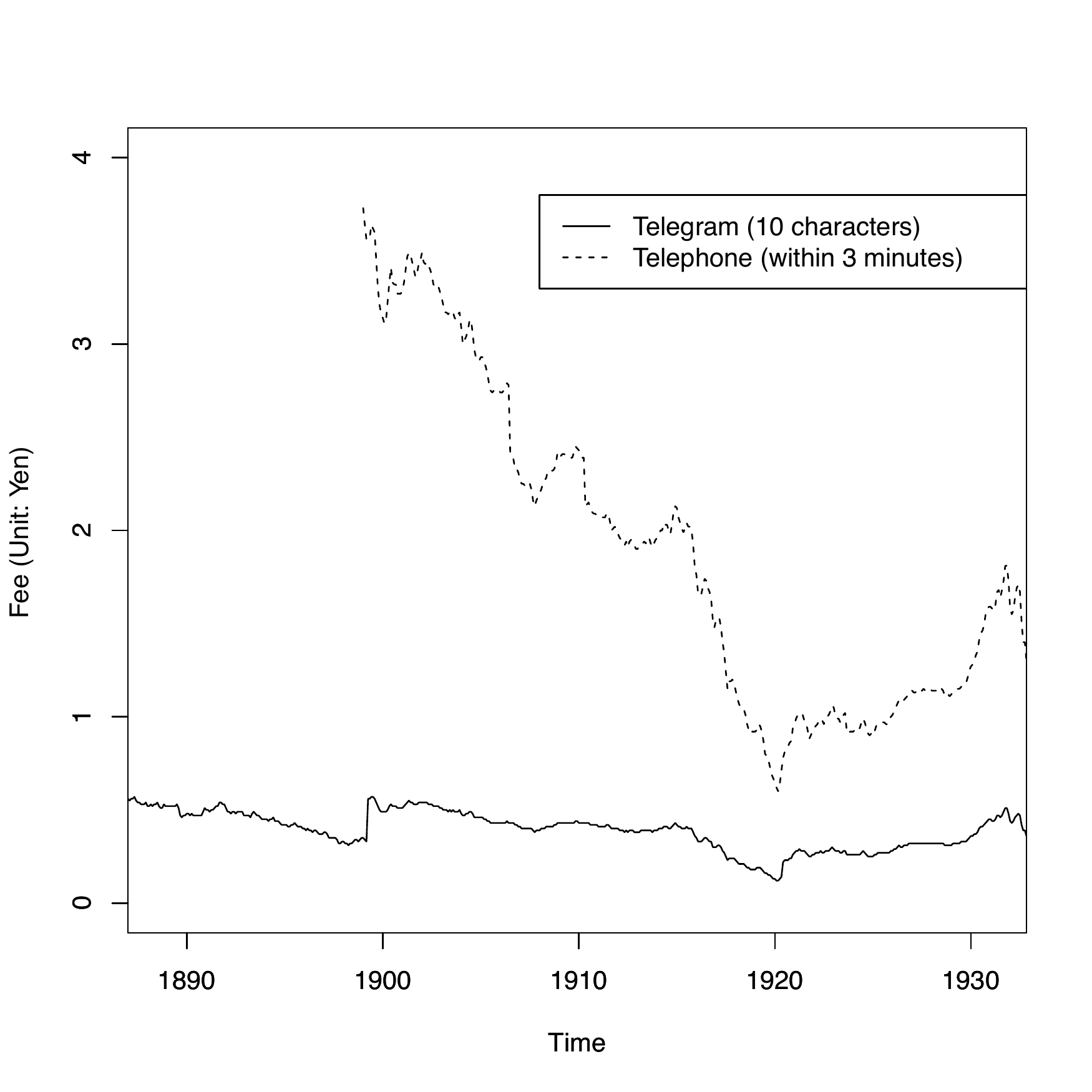}
\vspace*{3pt}
{
\begin{minipage}{350pt}
 \footnotesize
 \underline{Data Sources}: 
\begin{itemize}
 \item[(1)] \citet{ntt1955htf,ntt1956htf}.
 \item[(2)] \citet{boj1987hys}.
\end{itemize}
 \underline{Notes}:
 \begin{itemize}
 \item[(1)] Japanese government launched the long-distance call service between Tokyo and Osaka in January 1899.
 \item[(2)] Telegram fee includes the additional charge of 10 characters for the senders' signature from March 1899 to May 1920 and the addressee fee after June 1920.
\end{itemize}
\end{minipage}}%
\end{center}
\end{figure}

We have defined the speed of market integration from the loading matrix of a TV-VEC model to examine to what extent markets are integrated. Figure \ref{mi_fig1} exhibits two panels: the time-varying speed of market integration for the Tokyo and Osaka rice exchanges, obtained from our estimates of the TV-VEC model, and corresponding series of its difference at the each period. The bottom panel of Figure \ref{mi_fig1} indicates that there are few periods in which the speed of market integration slowed down.
\begin{center}
(Figure \ref{mi_fig1} around here)
\end{center}
Figure \ref{mi_fig2} shows that the real fees of the telegram and telephone network between Tokyo and Osaka from 1887 to 1932, respectively.\footnote{We should use the consumer price index (CPI) to deflate the fees; however, the monthly CPI does not exist in our sample period. Therefore, we deflate the fees using the average of the wholesale price index (WPI) from 1934 to 1936. Note that the correlation coefficient between the annual average estimated CPI (which is calculated by \citet{okawa1967elt}) and WPI is very high (0.977) from 1887 to 1932.} According to the figure, although the real telephone fee fell greatly after the 1900s, the real telegram fee was lower than the real telephone fee during this period, which reflects the difference in charging systems between the telegram and telephone; the telegram fee was independent of the distance between users and telephone fee depended on it. Specifically, calls between users within about 200 kilometers (about 124 miles) of one another occupied 81\% of the long distance calls in 1932. Moreover, long distance calls were mainly used for short- and medium-distance communication. On the other hand, telegrams between users more than 200 kilometers from one another occupied 80\% of all telegrams in 1932, as they were mainly used for long distance communication (see \citet[pp.103--104]{kosaki1934mut}). According to the distance chart of telegraph, the length of telegraph cable between Tokyo and Osaka was 554 kilometers (344 miles). Since the distance was over 200 kilometers, the telegram was mainly used between the two cities.
\begin{center}
(Figure \ref{mi_fig2} around here)
\end{center}

\begin{figure}[bp]
 \caption{Sent Telegrams in Tokyo and Osaka}\label{mi_fig3}
 \begin{center}
 \includegraphics[scale=0.8]{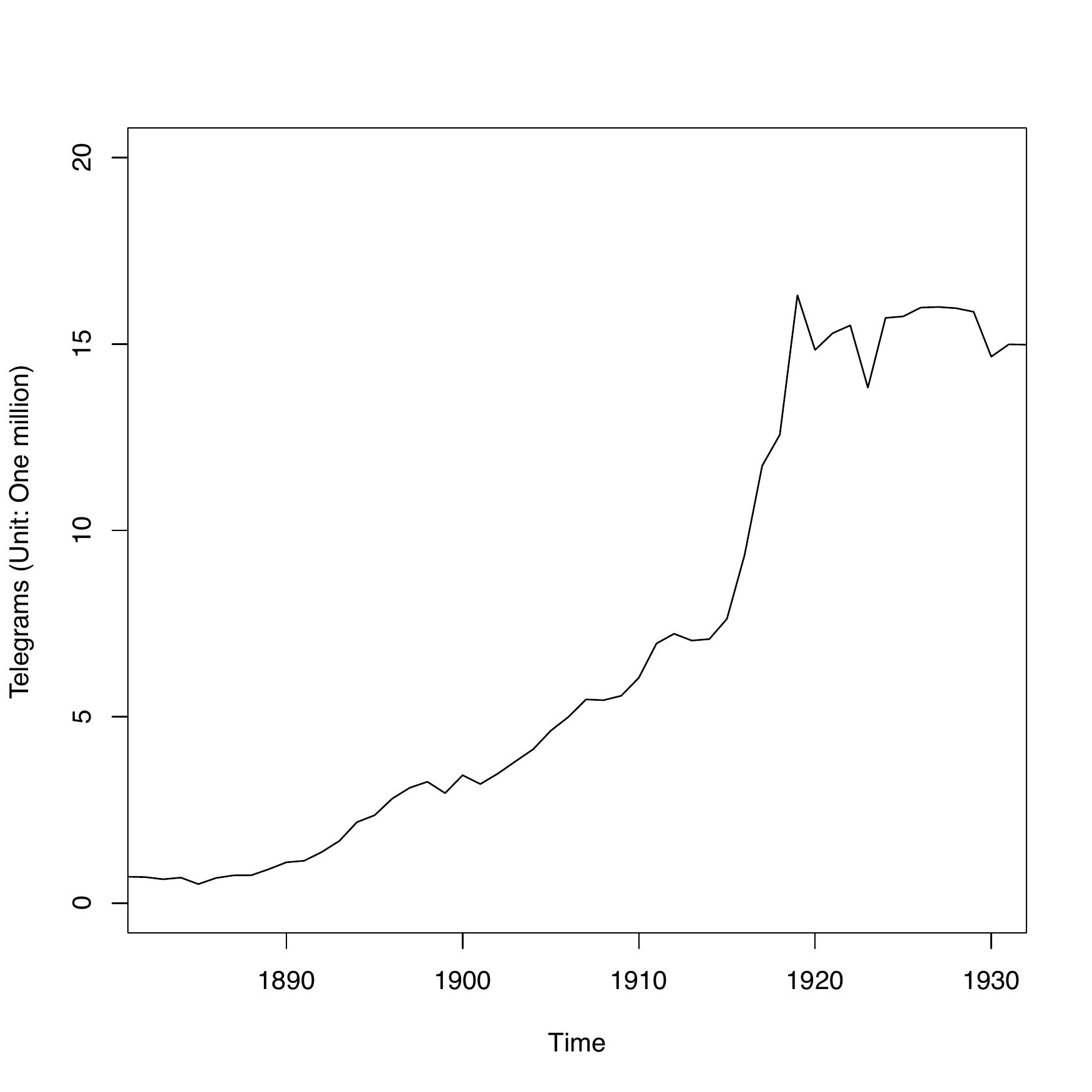}
\vspace*{3pt}
{
\begin{minipage}{350pt}
 \footnotesize
 \underline{Data Sources}: 
 \begin{itemize}
 \item[(1)] \citet{cabinet1883jsy,cabinet1884jsy}
 \item[(2)] \citet{cabinet1885jsy,cabinet1886jsy,cabinet1888jsy,cabinet1889jsy,cabinet1890jsy,cabinet1891jsy,cabinet1892jsy,cabinet1898jsy,cabinet1899jsy,cabinet1900jsy,cabinet1901jsy,cabinet1902jsy,cabinet1903jsy,cabinet1904jsy,cabinet1905jsy,cabinet1906jsy,cabinet1907jsy,cabinet1908jsy,cabinet1909jsy,cabinet1910jsy,cabinet1911jsy,cabinet1914jsya,cabinet1914jsyb,cabinet1915jsy,cabinet1916jsy,cabinet1918jsy,cabinet1919jsy,cabinet1922jsy,cabinet1925jsy,cabinet1926jsy,cabinet1927jsy,cabinet1928jsy,cabinet1929jsy,cabinet1931jsy,cabinet1932jsy,cabinet1933jsy,cabinet1934jsy}.
 \item[(3)] \citet{cabinet1893jsy,cabinet1894jsy,cabinet1895jsy,cabinet1896jsy,cabinet1897jsy}.
 \item[(4)] \citet{census1920jsy,census1921jsya,census1921jsyb}.	    
 \item[(5)] \citet{osaka1893ops,osaka1922ops,osaka1927ops}.
 \item[(6)] \citet{tokyo1893tps,tokyo1923tps,tokyo1928tps}.
 \end{itemize}
\end{minipage}}%
  \end{center}
\end{figure}

Based on the abovementioned information, we discuss the number of telegrams that were sent in Tokyo and Osaka from 1881 to 1932 (as shown in Figure \ref{mi_fig3}). We are supposed to discuss the number of telegrams exchanged between the two cities, but there are no available statistics that present such data. Accordingly, we discuss the data shown in Figure \ref{mi_fig3} as a proxy for the number of telegrams exchanged between Tokyo and Osaka. In fact, Tokyo had the largest share of sent telegrams from Osaka, and Osaka had the second-largest share of sent telegrams from Tokyo in the late 1930s (\citet[pp.743--744]{mct1940htb3}).
\begin{center}
(Figure \ref{mi_fig3} around here)
\end{center}

The trend of sending telegrams shown in Figure \ref{mi_fig3} exhibits a similar shape to the time-varying speed of market integration between Tokyo and Osaka shown in Figure \ref{mi_fig1}. In particular, we confirm the robustness of the interrelationship between the annual average time-varying speed of market integration from 1881 to 1932 is computed and annual sending telegrams.\footnote{There does not exist monthly sending telegrams. So, we creat annual average time-varying speed of market integration from monthly data.} We test a long-run relationship between the two variables and estimate a standard VEC model. Table \ref{mi_table5} presents the results of the cointegration tests and a standard VEC estimation. Both the tests reject the null hypothesis of ``$r=0$ (no cointegration),'' in favor of ``$r=1$ (atmost one cointegration)`` at the 1\% significance level. This indicates the presence of one cointegrating relationships. 
\begin{center}
(Table \ref{mi_table5} around here)
\end{center}
We assume a standard VEC model for our testing the robustness. Then, we regress the difference of the annual average time-varying speed of market integration and annual sending telegrams on their lagged values and error correction terms of a standard VEC model. Table \ref{mi_table5} summarizes our testing the robustness. Notice that the all error correction terms significantly affect each other. This is consistent with the above results of the cointegration tests.

Along with the growing telecommunications between the two cities, activated mainly by appreciation in the number of sent telegrams, the speed of market integration between Tokyo and Osaka significantly increased. Since the evolution of communication ameliorates efficiency in commerce, this correspondence seems quite natural. However, the trend of Figure \ref{mi_fig1} differs from that of Figure \ref{mi_fig3} in two points. First, while the speed of market integration between Tokyo and Osaka notably increased in the 1880s, the appreciation in the number of sent telegrams only moderately increased. Second, while the increase in the speed of market integration between Tokyo and Osaka lessened in the 1900s, the number of sent telegrams was continuously increasing. Consequently, we will mainly discuss what changed the speed of market integration between the two cities, focusing on these two points.


\subsubsection{Improvement of Telegraphic Technique and Diffusion of Press in 1880s}

Whereas the number of sent telegrams gently increased until the 1880s, the following three factors caused the rapid increase in the speed of market integration between Tokyo and Osaka (see Figures \ref{mi_fig1} and \ref{mi_fig3}). First, the required time for telegraph communication shortened. In 1882, the Telegraph Bureau of the Ministry of Engineering in Japan introduced the Wheatstone Automatic Telegraphic Machine from the UK. This machine could transmit and receive 400 characters in one minute, which greatly shortened connection time (see \citet[p.337]{mct1940htb3}).

Second, the following refinement of equipment improved the quality of telegraphic communication. Until the beginning of the 1880s, the Telegraph Bureau laid the telegraphic wire with the aerial insulator imported from the UK. However, since its insulator was unsuitable for the humid climate in Japan, it often short-circuited; as such, Japanese telegraph users faced poor communication quality. Accordingly, the Telegraph Bureau began to employ the Japanese advanced insulator for the telegraphic wire between Tokyo and Nagasaki, and the quality of telegraphic communication between Tokyo and Osaka was improved (see \citet[p.441--444]{mct1940htb3}).

Third, more newspapers and business magazines were published in the end of the 19th century than had been before. Publishers had frequently used telegrams to obtain market information and distributed the price data among merchants and businesspersons. Such publishers, located in the cities, especially in Tokyo and Osaka, issued their initial numbers around 1880. For example, ``{\it{Jiji Shimpo}}'' (Jiji Newspaper) in Tokyo and ``{\it{Asahi Shimbun}}'' (Asahi Newspaper) in Osaka made their first appearances in 1882 and 1879, respectively. Moreover, ``{\it{Tokyo Keizai Zasshi}}'' (Tokyo Economic Magazine), a well-known business magazine in prewar Japan, was launched in 1879 (see \citet[p.224]{sugihara1990tbm}). As a result, from 1878 to 1889, copies of newspapers and magazines increased from 31 to 102 million (see \citet[p.441]{cabinet1883jsy}; \citet[p.531]{cabinet1891jsy}). Even merchants who did not use the telegram were able to gain the latest market information in remote locations because of the diffusion of newspapers and magazines.

In the 1890s, the speed of market integration between the two cities increased along with the growth of sent telegrams (see Figures \ref{mi_fig1} and \ref{mi_fig3}). The Ministry of Communication set the telegram fee between Tokyo and Osaka at 0.15 yen in July 1885, and left it unchanged until February 1899 (see Figure \ref{mi_fig2}).\footnote{In 1885, the Telegraph Bureau was combined with the Postal Bureau, Mercantile Marine Bureau of the Ministry of Agriculture and Commerce, and Lighthouse Bureau of the Ministry of Engineering into the Ministry of Communications.} However, the consumer price index increased by almost 1.3 times in the same period (see \citet[p.135]{okawa1967elt}). Therefore, the real telegram fee was reduced, while the number of sent telegrams increased.

\subsubsection{Rising Telegram Fee and Interruption of Increase in Telephone Subscribers from the 1890s to the Mid-1900s}

The increase in the speed of market integration between Tokyo and Osaka became moderate from the end of the 1890s to mid-1900s (see Figure \ref{mi_fig1}), which was the result of two different points. The first point involves the decrease in the appreciation rate of the speed of market integration from the end of the 1890s to early 1900s, and the second point involves the similar situation from the early to mid-1900s. Accordingly, we will discuss these two points separately.

The first point was an interruption of the increase in sent telegrams from 1899 to 1902 (see Figure \ref{mi_fig3}). In short, the decrease in the appreciation rate of the speed of market integration around 1900 is explainable in terms of the trend of sending telegrams. In April 1899, the Ministry of Communications raised the telegram fee (see Figure \ref{mi_fig2}). From then to 1902, the Japanese economy slipped into stagnation. In fact, in that time, the real GDP of Japan grew slightly from 6.3 to 6.4 million yen (see \citet[p.213]{okawa1974elt}; \citet[p.60]{nishikawa1990jeh}). During the recession, the consumer price index slightly increased by 1.2 percent from 1900 to 1902 (see \citet[p.135]{okawa1967elt}). Therefore, the real telegram fee did not decrease until 1902 (see Figure \ref{mi_fig1}). However, the Japanese economy went through a phase of recovery in 1903, and commodity prices began to increase. As such, because of the downward trend of the real telegram fee, the number of sent telegrams increased again.

\begin{figure}[bp]
 \caption{Local Telephone Calls in Tokyo and Osaka}\label{mi_fig4}
 \begin{center}
 \includegraphics[scale=0.8]{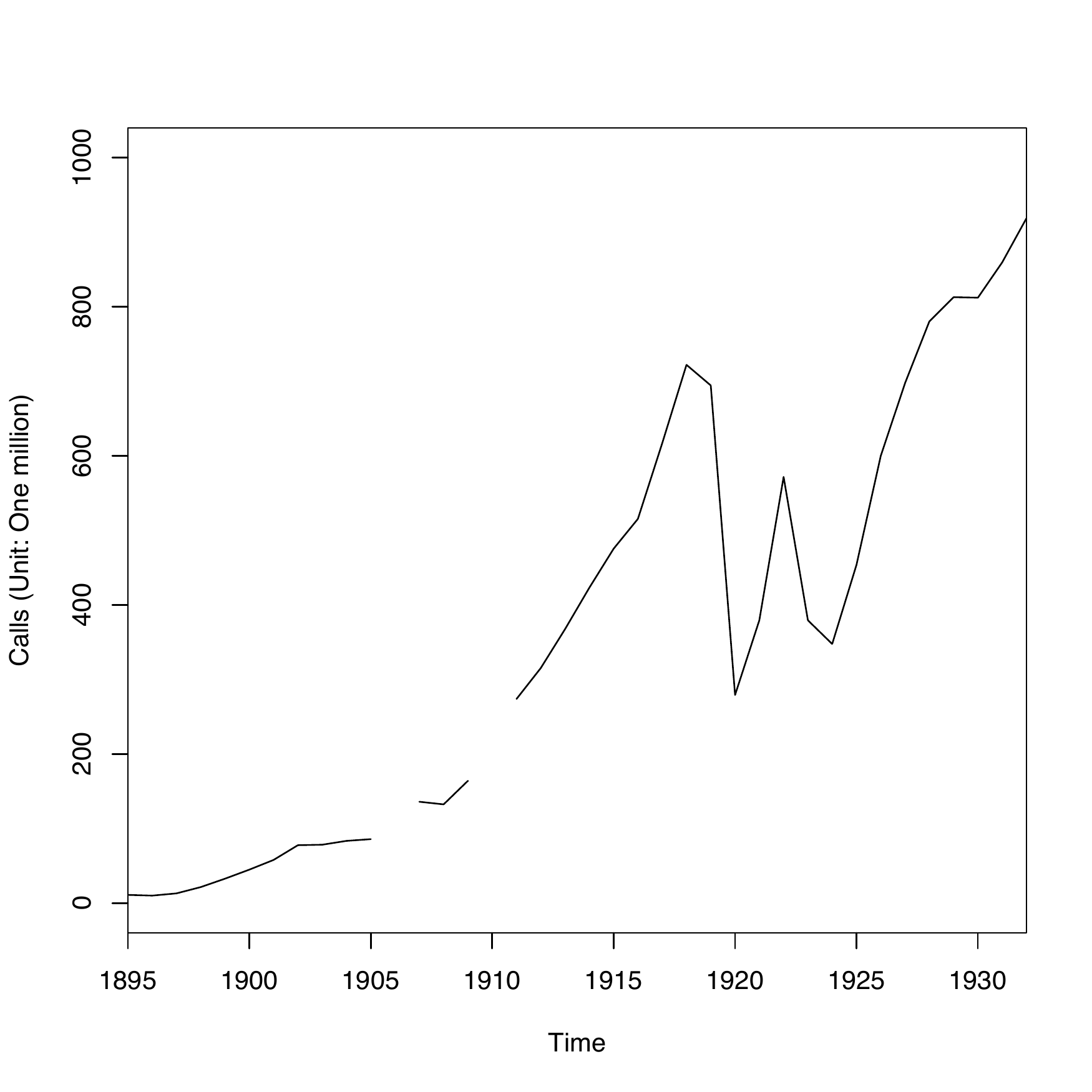}
\vspace*{3pt}
{
\begin{minipage}{350pt}
 \footnotesize
 \underline{Data Sources}: 
\begin{itemize}
 \item[(1)] \citet{cabinet1898jsy,cabinet1899jsy,cabinet1900jsy,cabinet1901jsy,cabinet1902jsy,cabinet1903jsy,cabinet1904jsy,cabinet1905jsy,cabinet1906jsy,cabinet1907jsy,cabinet1908jsy,cabinet1910jsy,cabinet1914jsya,cabinet1914jsyb,cabinet1915jsy,cabinet1916jsy,cabinet1918jsy,cabinet1919jsy,cabinet1922jsy}.
 \item[(2)] \citet{cabinet1896jsy,cabinet1897jsy}.
 \item[(3)] \citet{tokyo1908tps,tokyo1913tps,tokyo1934tps}.
 \item[(4)] \citet{osaka1923ops,osaka1925ops,osaka1932ops,osaka1936ops}.
\end{itemize}
 \underline{Note}: There is no statistics presenting the number of units on the local telephone call in 1906 and 1910.
\end{minipage}}%
\end{center}
\end{figure}

The second point was another interruption of the increase in units on the local telephone calls in Tokyo and Osaka from 1903 to 1908 (see Figure \ref{mi_fig4}). The decrease in the appreciation rate of the speed of market integration in the mid01900s is explainable not in terms of the trend of sending telegrams but in terms of the interruption of increase in the use of the service with local telephone as mentioned below. In 1890, the Ministry of Communications launched a new service to shorten the required time for sending and receiving telegrams. In the same year, the telegraph office began to send and receive the content of telegrams through local telephone service, acting as a hybrid form of telecommunication between the local telephone and remote telegram services for local telephone subscribers around the same time. The subscribers had to pay 66 yen as the annual telephone charge, which was approximately 440 times higher than the telegram fee between Tokyo and Osaka. However, they could use the local telephone service without an additional charge. As a result, local telephone calls in Tokyo and Osaka continuously increased until 1919 (see Figure \ref{mi_fig4}).
\begin{center}
(Figure \ref{mi_fig4} around here)
\end{center}

\begin{figure}[bp]
 \caption{Local Telephone Subscribers in Tokyo and Osaka}\label{mi_fig5}
 \begin{center}
 \includegraphics[scale=0.8]{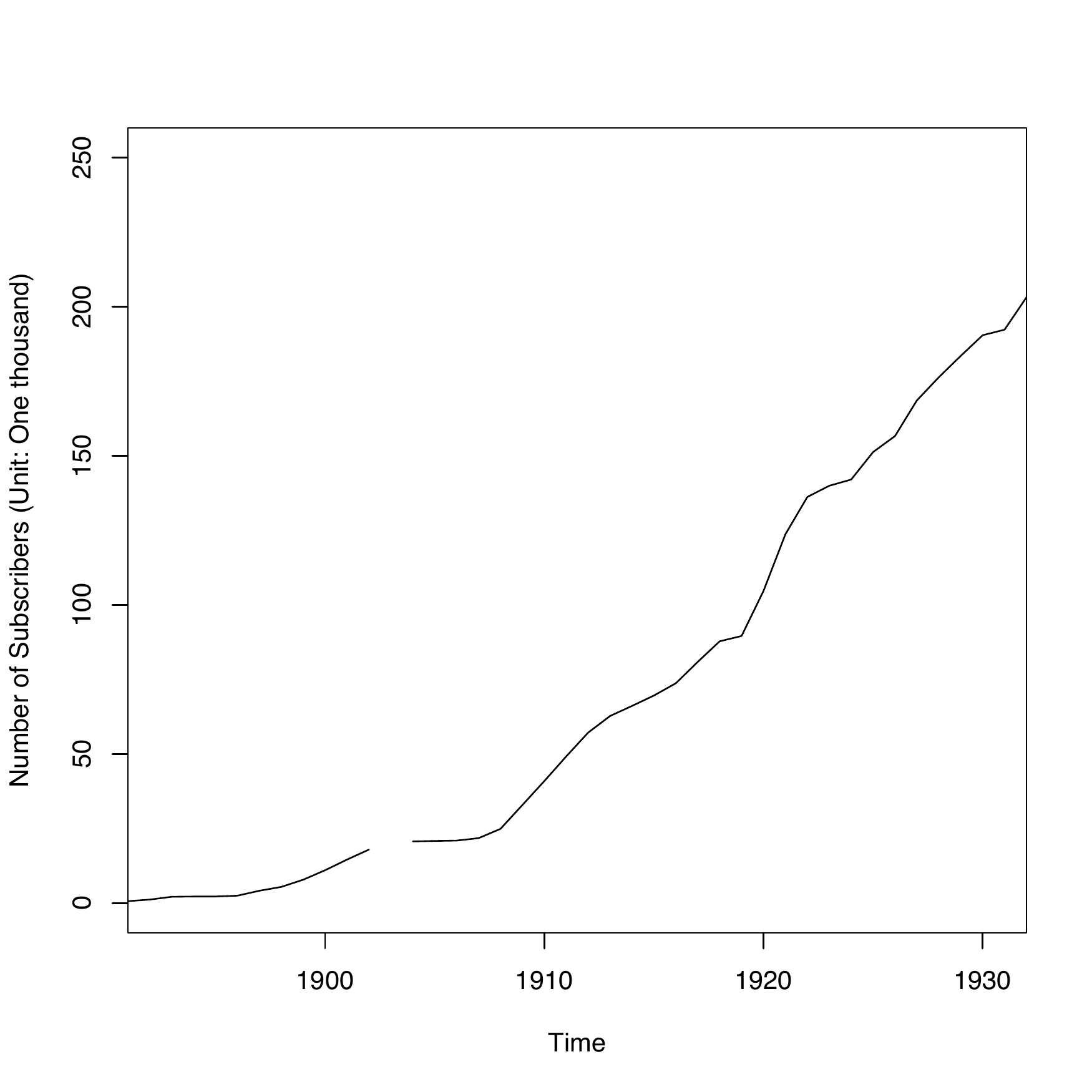}
 \vspace*{3pt}
{
\begin{minipage}{350pt}
 \footnotesize
 \underline{Data Sources}: 
\begin{itemize}
 \item[(1)] \citet{cabinet1892jsy,cabinet1898jsy,cabinet1899jsy,cabinet1900jsy,cabinet1901jsy,cabinet1902jsy,cabinet1903jsy,cabinet1904jsy,cabinet1905jsy,cabinet1906jsy,cabinet1907jsy,cabinet1908jsy,cabinet1910jsy,cabinet1911jsy,cabinet1914jsya,cabinet1914jsyb,cabinet1915jsy,cabinet1916jsy,cabinet1918jsy,cabinet1919jsy}.
 \item[(2)] \citet{cabinet1893jsy,cabinet1894jsy,cabinet1895jsy,cabinet1896jsy,cabinet1897jsy}.
 \item[(3)] \citet{census1920jsy,census1921jsya}.	    
 \item[(4)] \citet{osaka1922ops,osaka1924ops,osaka1927ops,osaka1932ops,osaka1936ops}.
 \item[(5)] \citet{tokyo1927tps,tokyo1934tps}.
\end{itemize}
 \underline{Note}: There is no statistics presenting the total number of the local telephone subscribers in 1903.
\end{minipage}}
\end{center}
\end{figure}

This increase in the local telephone calls resulted from the growth of the telephone subscribers (see Figure \ref{mi_fig5}).
\begin{center}
(Figure \ref{mi_fig5} around here)
\end{center}
We can examine these subscribers' jobs in Tokyo and Osaka according to detailed data from March 1897, in which there were 2487 telephone subscribers in Tokyo and Osaka. Then, 1124 subscribers were categorized into merchants, who needed to acquire the commercial information, and included 154 brokers in the exchanges, 123 cloth traders, 88 restaurants, and 70 traders of rice and manure. Therefore, many rice traders, including brokers in the exchange, wholesalers, and retailers, had been able to use the local telephone service frequently since the 1890s (see \citet[pp.856--857]{cabinet1889jsy}).

These telephone subscribers could enjoy the abovementioned service provided by telegraph offices, the procedure of which went roughly as follows. First, a sender provided the telegraph office with the content of a telegram by telephone. Second, the telegraph office gave a telegram to the telegraph office nearest the destination. Third, the receiving telegraph office provided an addressee with the content of the telegram over the telephone.

This service had two points of convenience for users. First, it reduced the required time for sending and receiving telegrams. A user sending a telegram did not need to go to the telegraph office or wait for a telegram carrier (see \citet[p.535]{mct1940htb3}). Second, this service simplified the payment of the telegram fee. Moreover, the fee for this service was free until November 1897. However, the Ministry of Communications switched from free to paid service in December 1897, and charged users 0.05 yen per telegram.\footnote{See \citet{asahi1897srt,asahi1899rsf} for details.} Yet, this service fee was reduced to 0.03 yen on January 15, 1899. In short, the service fee was generally low, and was charged together with the telegram fee at the end of each month. As a result, it was convenient for telephone subscribers of a merchant to frequently exchange commercial information using this service.\footnote{See \citet{asahi1900srt} for details.}

The statistics presenting the number of telegrams using the service were unavailable before the early 1920s because the internal documents of the Ministry of Communications were destroyed by a fire caused by the Great Kanto Earthquake in 1923 (see \citet[p.680]{ntt1959htt}). However, the telephone subscribers frequently used the service to send and receive telegrams. Moreover, \citet[p.318]{tct1958htc} stated, ``The use of the sending telegram service with local phone increased along with the growing telephone subscribers.'' In fact, the Tokyo Central Telegraph Office laid the dedicated telephone lines for the sending telegram service by local telephone in 1900, but the lines were overflowed due to more users than their capacity in 1909 (see \citet[p.263]{tct1958htc}). The Tokyo Central Telegraph Office was forced to change the dial number for sending telegrams to alleviate the situation in 1909.\footnote{See \citet{asahi1909tcd} for details.}

In summary, the sending telegram service with local telephone allowed telephone subscribers to exchange messages on markets among merchants quite quick and inexpensively. Therefore, this service accelerated the increase in the speed of market integration between Tokyo and Osaka. However, the expanding utilization of the service was interrupted in the mid-1900s along with the presence of stagnant telephone subscribers in Tokyo and Osaka (see Figure \ref{mi_fig5}).

The Ministry of Communications expanded telephone facilities after the Japanese-Sino War. Then, Japan aggravated diplomatic relations with Russia in 1903 and the Japanese-Russo War began in 1904. The Ministry of Communications suspended the expansion of telephone facilities for civilian use, giving priority to military use. Accordingly, because the increase in telephone subscribers was interrupted, local telephone calls in Tokyo and Osaka did not increase from 1903 to 1908. Therefore, the use of the sending telegram service with local phone did not increase, and the increase in the speed of market integration between Tokyo and Osaka lessened from the early to mid-1900s.

\subsubsection{Diffusion of Telegram and Local Telephone from the End of 1900s to the Mid-1910s}

The speed of market integration between the two cities increased from the end of the 1900s to mid-1910s (see Figure \ref{mi_fig1}). We discuss this situation from two viewpoints to understand from what it resulted. The first point was the increasing number of sent telegrams from the late 1900s (see Figure \ref{mi_fig3}). Specifically, the sending of telegrams rapidly increased because the real telegram fee was reduced in the late 1910s (see Figure \ref{mi_fig2} and \ref{mi_fig3}). The Ministry of Communications had not changed the telegram fee from April 1899 to May 1920, while the consumer price index increased by 3.3 times during the same period (see \citet[p.135]{okawa1967elt}). Consequently, the real telegram fee was reduced.

he second point was the increasing number of local telephone calls from 1909 (see Figure \ref{mi_fig4}). In 1907, the Ministry of Communications resumed the expansion of telephone facilities for civilian use, and the number of telephone subscribers increased in Tokyo and Osaka (see \citet[pp.555-560]{mct1940htb4}, \citet[pp.71--72]{fujii1998edt}, and Figure \ref{mi_fig5}). As a result, local telephone calls in Tokyo and Osaka increased, and more telephone subscribers used the sending telegram service with local telephone.

\subsubsection{Change in Telephone Charge System and Great Kanto Earthquake from End of 1910s}

\begin{figure}[bp]
 \caption{Long-Distance Telephone Calls in Tokyo and Osaka}\label{mi_fig6}
 \begin{center}
 \includegraphics[scale=0.8]{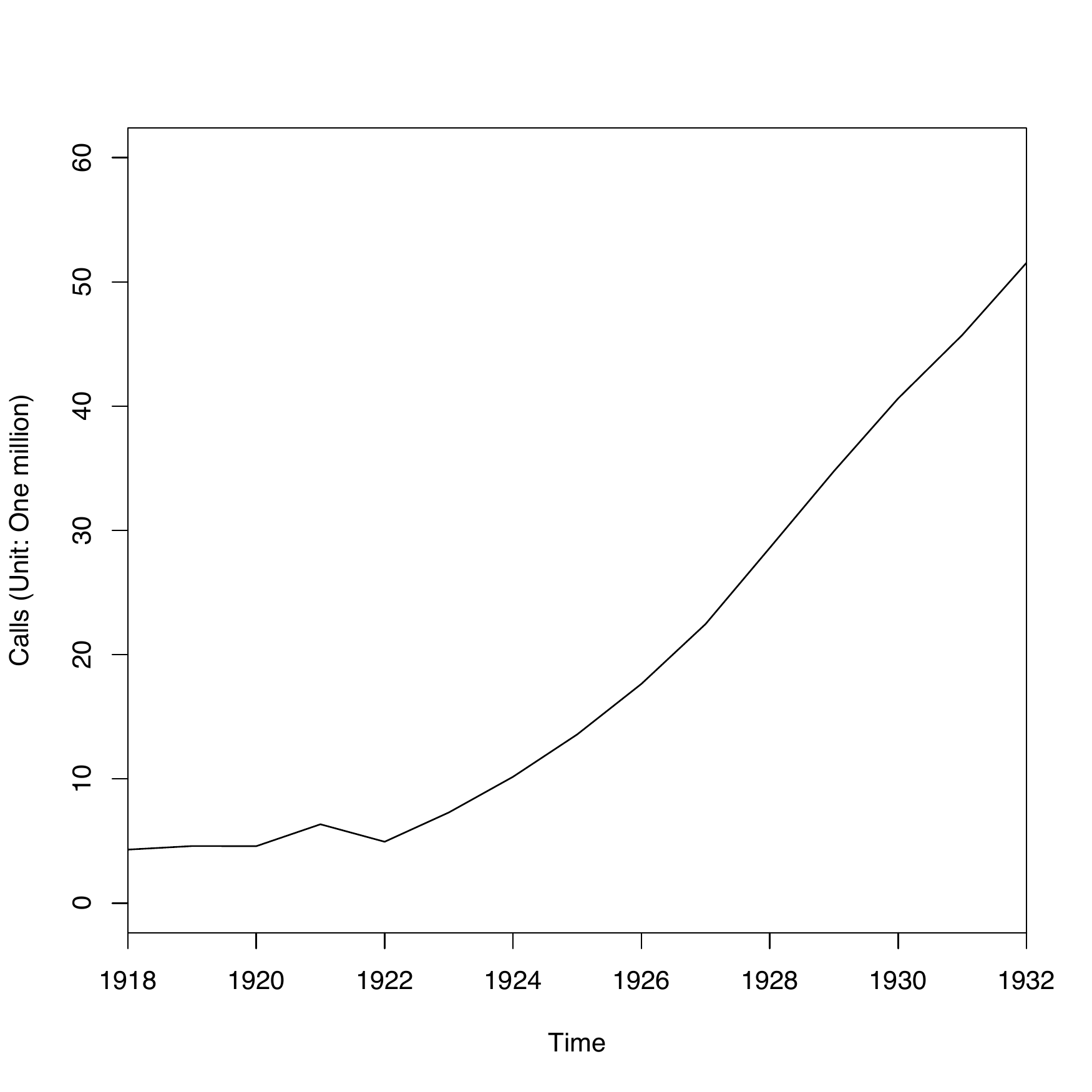}
\vspace*{3pt}
{
\begin{minipage}{350pt}
\footnotesize
\underline{Data Sources}: 
\begin{itemize}
 \item[(1)] \citet{osaka1924ops,osaka1929ops,osaka1933ops,osaka1936ops}.
 \item[(2)] \citet{tokyo1927tps,tokyo1934tps}.
\end{itemize}
\end{minipage}}%
\end{center}
\end{figure}

The increase in the speed of market integration between the two cities was suspended in the end of the 1910s, along with the interruption in the growth of sending telegrams (see Figure \ref{mi_fig1} and \ref{mi_fig3}). There were two points involved in the interruption: the rise in the real telegram fee in June 1920 (see Figure \ref{mi_fig2}) and expansion in the use of long-distance telephone calls in Tokyo and Osaka from the early 1920s (see Figure \ref{mi_fig6}).
\begin{center}
(Figure \ref{mi_fig6} around here)
\end{center}

The increase in the number of long-distance telephone calls resulted from the shortened call connection time. In 1924, the Ministry of Communications restricted users' time for long distance calls to three minutes to alleviate telephone network congestion (see \citet[p.304]{mct1940htb4}). At the same time, the long distance telephone service was a substitute for short- and medium-distance telegrams. When users telecommunicated within 200 kilometers (about 124 miles) of one another, the long distance telephone call cost them less than did a telegram (see \citet[p.128]{fujii1998edt}). However, according to Figure \ref{mi_fig3}, sending telegrams in Tokyo and Osaka did not decrease in the 1920s. These facts suggest that the number of sent long distance telegrams (over 200 kilometers) increased, while that of short and medium distance telegrams decreased. That is, the number of telegrams exchanged between Tokyo and Osaka was presumed to increase; nevertheless, the increase in the speed of market integration between the two cities was suspended. This situation resulted from the decrease in local telephone calls in Tokyo and Osaka (see Figure \ref{mi_fig4}).

The number of local telephone calls in the two cities decreased greatly because of a change in the telephone charge system in April 1920. Although telephone subscribers could use the local telephone without an additional fee, except for the annual charge until March 1920, they had to pay 0.02 yen for one call in large cities, such as Tokyo, Yokohama, Nagoya, Kyoto, Osaka, and Kobe (see \citet[pp.33--37]{mct1940htb4}). The Ministry of Communications changed the telephone charge system to alleviate network congestion, similar to the case of time restrictions of long distance calls in 1924. On the other hand, the number of telephone subscribers increased because the Ministry of Communications reduced the annual charge from 66 to 45 yen in April 1920, and local telephone calls increased from 1921 to 1922 (see Figures \ref{mi_fig4} and \ref{mi_fig5}). However, the use of local telephone calls decreased again because of the Great Kanto Earthquake in 1923, in which 64\% of the building owned by telephone subscribers in Tokyo was destroyed and 15 telegraph offices, including the Tokyo Central Telegraph Office, collapsed (see \citet[p.165]{mct1940htb3} and \citet[p.642]{mct1940htb4}). In short, local telephone calls decreased in the early 1920s because of the change in the charge system in 1920 and the devastating earthquake in 1923. Therefore, the use of the sending telegram service with local telephone decreased.

In the late 1920s, the speed of market integration between Tokyo and Osaka increased slightly (see Figure \ref{mi_fig1}). Local telephone calls began to increase in 1925 with the rebuilding after the earthquake in Tokyo and, in 1928, they exceeded the pre-earthquake level, which caused the expansion of the sending telegram service with local telephone. We can examine the utilization of this service within the district boundary of the Tokyo Central Telegraph Office, according to detailed data on the service from 1924. From 1924 to 1931, the use of this service increased from 117005 to 384989 (see \citet[p.319]{tct1958htc}). At the same time, local telephone calls in Tokyo increased from 161 to 568 million (see \citet[p.588]{tokyo1927tps} and \citet[p.564]{tokyo1934tps}). That is, from 1924 to 1931, in Tokyo, the sending telegrams with local telephone service increased 3.3 times, and local telephone calls increased 3.5 times. Therefore, the use of the sending telegrams with local phone service increased along with local telephone calls. As a result, the speed of market integration between the two cities increased.

\section{Conclusion}\label{mi_sec6}
The main results of this paper are summarized in three points: (i) the Japanese rice markets had been integrated until the 1910s, (ii) increasing use of telegraphs had accelerated rice market integration from the Meiji period in Japan and (iii) the local telephone service, which reduced urban users' time spent sending and receiving telegrams, promoted market integration.

The previous studies concluded that the Japanese rice markets were integrated between Osaka and other local cities in the Tokugawa period (see Section \ref{mi_sec1}). In contrast, this paper argues that the major rice markets had continued to be integrated until the 1910s because of continuously increasing use of telegrams and local telephone. The government reduced the real telegram fee and local telephone subscribers paid an annually fixed fee to use their telephones. Accordingly, brokers could use the telegram and local telephone services inexpensively. Meanwhile, the long distance telephone fee was higher than was the telegram fee. As a result, traders did not substitute the long distance telephone service for that of the telegram. In summary, the trend of telegraph use depended on the fee structure, and the reduction of the real telegram fee resulted in the growing use of telegrams. Therefore, the increase in the real fee caused market integration to stagnate.

As we mentioned in Section \ref{mi_sec1}, economic historians assert that the introduction of telegraph network was a fundamental factor of market integration from the early modern period. However, they do not focus on the change in use of telegraph. We argue that the growing use of telegraph promoted market integration after the introduction of telecommunication network. Particularly, in Japan, the process of market integration depended on the communication costs determined by the government. As we mentioned in Section \ref{mi_sec2}, the Japanese government had built and managed the whole set of telecommunication networks since the 1870s. Accordingly, it set the fees of telegrams and telephone calls. Thus, market integration in Japan strongly depended on the government's policy on telecommunication before the early 20th century. In contrast to the development process of telegraph in Japan, private firms often laid telegraph lines in Europe. Conversely, the government in Europe generally nationalized the telecommunication firms and obtained the authority to determine the fees of telegrams and telephone calls from the late 19th century (see \citet[p.19]{noam1992tie}). We conclude that the telecommunication policy affected the process of market integration in not only Japan but also Europe.

\section*{Acknowledgments}

We would like to thank Shigehiko Ioku, Kozo Kiyota, Kris Mitchener, Chiaki Moriguchi, Tetsuji Okazaki, Minoru Omameuda, Masato Shizume, Yasuo Takatsuki, Tatsuma Wada, and Asobu Yanagisawa for their helpful comments and suggestions. We would also like to thank seminar and conference participants at Keio University; the Japanese Economics Association 2016 Autumn Meeting; the 85th Annual Conference of the Socio-Economic History Society; and the 91th Annual Conference of the Western Economic Association International for helpful discussions. We also thank the Japan Society for the Promotion of Science for their financial assistance, as provided through the Grant in Aid for Scientific Research Nos.26380397 (Mikio Ito), 26780199 (Kiyotaka Maeda), and 15K03542 (Akihiko Noda). All data and programs used for this paper are available on request.

\setcounter{table}{0}
\renewcommand{\thetable}{\arabic{table}}

\clearpage

\clearpage

\clearpage

\begin{table}[tbp]
\caption{Descriptive Statistics and Unit Root Tests}\label{mi_table1}
\begin{center}
\scriptsize
\begin{tabular}{lllrrrrrrrrrr}\hline\hline
 &  &  & \multicolumn{4}{c}{Level} &  & \multicolumn{4}{c}{First Difference} & \\\cline{4-7}\cline{9-12}
 &  &  & \multicolumn{1}{c}{$S_T$} & \multicolumn{1}{c}{$F_T$} & \multicolumn{1}{c}{$S_O$} & \multicolumn{1}{c}{$F_O$} &  & \multicolumn{1}{c}{$\Delta S_T$} & \multicolumn{1}{c}{$\Delta F_T$} & \multicolumn{1}{c}{$\Delta S_O$} & \multicolumn{1}{c}{$\Delta F_O$} & \\\hline
 & Mean &  & 2.6811 & 2.6645 & 2.6648 & 2.6449 &  & 0.0009  & 0.0013  & 0.0013 & 0.0013 & \\
 & SD &  & 0.6328  & 0.6629  & 0.6629  & 0.6237  &  & 0.0539  & 0.0627  & 0.0619  & 0.0634  & \\
 & Min &  & 1.5282  & 1.4974 & 1.4469 & 1.5129 &  & $-0.3670$ & $-0.3189$ & $-0.2939$ & $-0.3069$ & \\
 & Max &  & 3.9947 & 3.9281 & 4.0157 & 3.9178 &  & 0.2791 & 0.3860 & 0.6569 & 0.2523 & \\\hline
 & ADF-GLS &  & $-2.1585$ & $-2.1695$ & $-2.0507$ & $-2.2208$ &  & $-3.8500$ & $-3.8161$ & $-13.8956$ & $-7.7518$ & \\
 & Lags &  & 1 & 1 & 1 & 1 &  & 8 & 8 & 1 & 3 & \\
 & $\hat\phi$ &  & 0.9926 & 0.9881 & 0.9890 & 0.9881 &  & 0.5372 & 0.3781 & 0.1968 & 0.3690 & \\\hline
 & $\mathcal{N}$ &  & \multicolumn{4}{c}{620} &  & \multicolumn{4}{c}{619} & \\\hline\hline
\end{tabular}
\vspace*{5pt}
{
\begin{minipage}{420pt}
\scriptsize
{\underline{Notes:}}
\begin{itemize}
\item[(1)] ``ADF-GLS'' denotes the ADF-GLS test statistics, ``Lags'' denotes the lag order selected by the MBIC, and ``$\hat\phi$'' denotes the coefficients vector in the GLS detrended series (see equation (6) in \citet{ng2001lls}).
\item[(2)] In computing the ADF-GLS test, a model with a time trend and a constant is assumed. The critical value at the 1\% significance level for the ADF-GLS test is ``$-3.42$''.
\item[(3)] ``$\mathcal{N}$'' denotes the number of observations.
\item[(4)] R version 3.4.1 was used to compute the statistics.
\end{itemize}
\end{minipage}}%
\end{center}
\end{table}

\clearpage

\begin{table}[tbp]
\caption{Cointegration Tests}\label{mi_table2}
  \begin{center}
\begin{tabular}{llrrrrrrrrl}\hline\hline
 & & & & & \multicolumn{2}{c}{Max Eigen} & & \multicolumn{2}{c}{Trace} & \\\cline{6-7}\cline{9-10}
 & & & Eigenvalues & & Test Stats & CV(1\%) & & Test Stats & CV(1\%)\\\hline
 & None & & 0.3587 & & 274.59 & 33.24 & & 433.69 & 60.16\\
 & At most 1 & & 0.1424 & & 94.94 & 26.81 & & 159.09 & 41.07\\
 & At most 2 & & 0.0958 & & 62.20 & 20.20 & & 64.16 & 24.60\\
 & At most 3 & & 0.0032 & & 1.96 & 12.97 & & 1.96 & 12.97\\\hline\hline
\end{tabular}
\vspace*{5pt}
{
\begin{minipage}{400pt}
\footnotesize
{\underline{Notes:}}
\begin{itemize}
 \item[(1)] ``Max Eigen'' and ``Trace'' denote the \citetapos{johansen1988sac} maximal eigenvalue test and \citetapos{johansen1991eht} trace test, respectively.
 \item[(2)] ``Test Stats'' and ``CV(1\%)'' denote the test statistics and the critical values at the 1\% significance level for the each tests, respectively.
 \item[(3)] R version 3.4.1 was used to compute the statistics.
\end{itemize}
\end{minipage}}%
  \end{center}
\end{table}

\clearpage

\begin{table}[tbp]
\caption{Time-Invariant VECM Estimations}\label{mi_table3}
\begin{center}
\begin{tabular}{lllccccc}\hline\hline
 &  &  & $\Delta S_{T,t}$ & $\Delta F_{T,t}$ & $\Delta S_{O,t}$ & $\Delta F_{O,t}$ & \\\hline
 & Difference &  &  &  &  &  & \\
 &  & \multirow{2}*{$\Delta{S_{T,t-1}}$} & 0.2476  & $-0.0482$  & 0.2890  & 0.0352  & \\
 &  &  & [0.0864]  & [0.1163]  & [0.1153]  & [0.0969]  & \\
 &  & \multirow{2}*{$\Delta{F_{T,t-1}}$} & 0.0783  & 0.1924  & 0.0497  & $-0.2737$  & \\
 &  &  & [0.0576]  & [0.0832]  & [0.0542]  & [0.0642]  & \\
 &  & \multirow{2}*{$\Delta{S_{O,t-1}}$} & 0.0069  & $-0.0677$  & $-0.0753$  & $-0.0365$  & \\
 &  &  & [0.0817]  & [0.0748]  & [0.0875]  & [0.0696]  & \\
 &  & \multirow{2}*{$\Delta{F_{O,t-1}}$} & 0.0153  & 0.0629  & 0.0140  & 0.2681  & \\
 &  &  & [0.0425]  & [0.0583]  & [0.0438]  & [0.0476]  & \\\hline
 & Level &  &  &  &  &  & \\
 &  & \multirow{2}*{Constant} & 0.0070  & 0.0234  & -0.0401  & 0.0148  & \\
 &  &  & [0.0111] & [0.0128] & [0.0152] & [0.0152] & \\
 &  & \multirow{2}*{$S_{T,t-1}$} & -0.1572  & -0.0672  & 0.1443  & -0.1555  & \\
 &  &  & [0.0536] & [0.0548] & [0.0942] & [0.0437] & \\
 &  & \multirow{2}*{$F_{T,t-1}$} & 0.0883  & -0.0540  & 0.0939  & 0.5389  & \\
 &  &  & [0.0506] & [0.0726] & [0.0479] & [0.0628] & \\
 &  & \multirow{2}*{$S_{O,t-1}$} & 0.0065  & 0.0510  & -0.3149  & 0.1238  & \\
 &  &  & [0.0467] & [0.0490] & [0.0796] & [0.0506] & \\
 &  & \multirow{2}*{$F_{O,t-1}$} & 0.0615  & 0.0628  & 0.0921  & -0.5150  & \\
 &  &  & [0.0355] & [0.0550] & [0.0371] & [0.0556] & \\\hline
 & ${\bar{R}}^2$ &  & 0.1792  & 0.0242  & 0.1836  & 0.2835  & \\
 & $L_C$ &  &\multicolumn{4}{c}{93.3609} & \\\hline\hline
 \end{tabular}
\vspace*{5pt}
{
\begin{minipage}{350pt}
\footnotesize
{\underline{Notes:}}
\begin{itemize}
\item[(1)] ``${\bar{R}}^2$'' denotes the adjusted $R^2$, and ``$L_C$'' denotes \citeapos{hansen1992a}{hansen1992a,hansen1992b} joint $L_C$ statistic with variance.
\item[(2)] \citetapos{newey1987sps} robust standard errors are in brackets.
\item[(3)] R version 3.4.1 was used to compute the estimates and the statistics.
\end{itemize}
\end{minipage}}%
  \end{center}
\end{table}

\clearpage

\begin{table}[tbp]
\caption{Cointegration Order Change Tests}\label{mi_table4}
 \begin{center}
\begin{tabular}{cccccccc}\hline\hline
& & & $SupQ^1$ & $SupQ^2$ & $WQ$ & $SQ$ &\\\cline{4-7}
& Test Statistics & & 0.59 & 0.85 & 0.59 & 1.14 &\\
& CV (1\%) & & 4.49 & 5.91 & 4.96 & 7.67 &\\\hline\hline
\end{tabular}
{
\begin{minipage}{250pt}
\vspace*{3pt}\scriptsize
{\underline{Notes:}}
\begin{itemize}
 \item[(1)] ``$SupQ^1$'' and ``$SupQ^2$'' denote the values allowing for one break and two breaks, respectively.
 \item[(2)] ``$WQ$'' and ``$SQ$'' also denote consistent test stats when we suppose less than three breaks based on the maximum and sum of $SupQ^1$ and $SupQ^2$, respectively.
 \item[(3)] R version 3.4.1 was used to compute the statistics.
\end{itemize}
\end{minipage}}%
\end{center}
\end{table}

\clearpage

\clearpage

\clearpage

\clearpage

\begin{table}[tbp]
\caption{Testing the Robustness}\label{mi_table5}
  \begin{center}
\begin{tabular}{lllccl}\hline\hline
 &  & \multicolumn{1}{c}{Eigenvalues} & Max Eigen & Trace & \\\hline
 & \multirow{2}*{None} & \multicolumn{1}{c}{\multirow{2}*{$0.5340$}} & $38.17$ & $48.61$ & \\
 &  &  & ($20.20$) & ($24.60$)\\
 & \multirow{2}*{At most 1} & \multicolumn{1}{c}{\multirow{2}*{$0.1884$}} & $10.43$ & $10.43$ & \\
 &  &  & ($12.97$) & ($12.97$)\\\hline\hline
 & & & & & \\\hline\hline
 &  &  & $\Delta\log(\zeta_t)$ & $\Delta \log(ST_t)$ & \\\hline
 & Difference &  &  &  & \\
 &  & \multirow{2}*{$\Delta\log(\zeta_{t-1})$} & $0.6184$  & $1.7351$  & \\
 &  &  & [$0.0782$]  & [$1.7610$]  & \\
 &  & \multirow{2}*{$\Delta\log(ST_{t-1})$} & $0.0043$  &$ -0.1587$  & \\
 &  &  & [$0.0072$]  & [$0.1111$]  & \\\hline
 & Level &  &  &  & \\
 &  & \multirow{2}*{Constant} & $-0.1010$ & $3.9223$  & \\
 &  &  & [$0.0464$]  & [$1.2666$]  & \\
 &  & \multirow{2}*{$\log(\zeta_t)$} & $-0.0576$  & $1.2199$  & \\
 &  &  & [$0.0100$]  & [$0.3561$]  & \\
 &  & \multirow{2}*{$\log(ST_{t-1})$} & $0.0079$  & $-0.2712$  & \\
 &  &  & [$0.0031$]  & [$0.0864$]  & \\\hline
 & $\bar{R}^2$ &  & $0.9349$  & $0.5001$  & \\
 & $\mathcal{N}$ &  & $50$  & $50$  & \\\hline\hline
\end{tabular}{
\begin{minipage}{400pt}
\vspace*{7pt}\footnotesize
\underline{Notes}:
\begin{itemize}
 \item[(1)] Top and bottom panels present the results of cointegration tests and the results of VEC estimations, respectively.
 \item[(2)] ``Max Eigen'' and ``Trace'' denote the \citetapos{johansen1988sac} maximal eigenvalue test and \citetapos{johansen1991eht} trace test, respectively.
 \item[(3)] Critical values at the 1\% significance level for the cointegration tests are in parentheses.
 \item[(4)] ``$\log(\zeta)$'' and ``$\log(ST)$'' denote the natural log of the annual average time-varying speed of market integration and annual sending telegrams, respectively.
 \item[(5)] \citetapos{newey1987sps} robust standard errors are in brackets.
 \item[(6)] R version 3.4.1 was used to compute the statistics.
\end{itemize}
\end{minipage}}%
\end{center}
\end{table}

\clearpage

\clearpage

\clearpage

\end{document}